\documentclass[aps,prd,twocolumn,showpacs,nofootinbib]{revtex4-1}
\usepackage{graphicx,hyperref}
\newcommand{\apjs}{ApJS}
\newcommand{\mnras}{MNRAS}
\newcommand{\wx}{w_{\rm X}}
\newcommand{\wy}{w_{\rm Y}}
\newcommand{\wz}{w_{\rm Z}}
\begin{document}

%=====================================================================
\title{New pulsar limit on local Lorentz invariance violation of
  gravity \\ in the standard-model extension}
\author{Lijing Shao}
\email{lshao@pku.edu.cn}
\affiliation{School of Physics, Peking University, Beijing 100871,
  China}   
\date{\today}
%=====================================================================
\begin{abstract}
  In the pure-gravity sector of the minimal standard-model extension,
  nine Lorentz-violating coefficients of a vacuum-condensed tensor
  field describe dominant observable deviations from general
  relativity, out of which eight were already severely constrained by
  precision experiments with lunar laser ranging, atom interferometry,
  and pulsars. However, the time-time component of the tensor field,
  $\bar s^{\rm TT}$, dose not enter into these experiments, and was
  only very recently constrained by Gravity Probe B. Here we propose a
  novel idea of using the Lorentz boost between different frames to
  mix different components of the tensor field, and thereby obtain a
  stringent limit of $\bar s^{\rm TT}$ from binary pulsars.  We
  perform various tests with the state-of-the-art white dwarf optical
  spectroscopy and pulsar radio timing observations, in order to get
  new robust limits of $\bar s^{\rm TT}$. With the isotropic cosmic
  microwave background as a preferred frame, we get $|\bar s^{\rm TT}|
  < 1.6 \times 10^{-5}$ (95\% CL), and without assuming the existence
  of a preferred frame, we get $|\bar s^{\rm TT}| < 2.8 \times
  10^{-4}$ (95\% CL). These two limits are respectively about 500
  times and 30 times better than the current best limit.
\end{abstract}
%=====================================================================
\maketitle

%=====================================================================
\section{Introduction}\label{sec:intro}
%=====================================================================

Einstein's general relativity (GR) has changed our understanding of
gravitation and spacetime for almost one hundred years. The success of
GR bases on its theoretical beauty and deep insights~\cite{mtw73}, as
well as its remarkable accuracy in explaining and predicting
experimental phenomena~\cite{will14}. The three classical tests
proposed by Einstein~\cite{ein16}, namely, i) the perihelion
precession of Mercury's orbit, ii) the deflection of light by the Sun,
iii) the gravitational redshift of light, established GR as the most
promising alternative for Newton's gravity theory. Later in 1960s,
tests of the Shapiro delay with the transmission of radar
pulses~\cite{sha64}, and astronomical discoveries of
quasars~\cite{ms63}, pulsars~\cite{hbp+68}, and cosmic microwave
background (CMB)~\cite{pw65}, reinforced GR's empirical foundation and
significance in various regimes. Subsequently, the systematic,
worldwide efforts after 1960s in tests of gravity have verified GR to
high precision~\cite{will14}.

Today, more aspects of gravitation are being explored. For instance,
the Gravity Probe B (GPB) have determined the geodetic and frame
dragging effects in GR to 0.3\% and 19\% respectively, by means of
cryogenic gyroscopes in Earth orbit~\cite{edp+11}. In a second
example, with the technique of timing with giant radio telescopes, the
Double Pulsar has verified GR to 0.05\%, and GR passed five tests
simultaneously in {\it one} system~\cite{ksm+06,bkk+08}. Needless to
say, we are going to witness the discovery of gravitational waves
(GWs) very soon with the global efforts from the GW communities. With
the new developments of ground-based and space-based laser
interferometric GW observatories~\cite{ss09,prrh11,gvlb13} and pulsar
timing arrays (PTAs)~\cite{hob13,kc13,mcl13}, a new era of
multi-wavelength, multi-message GW astronomy will soon open novel
possibilities to test the foundations of GR, especially to deeply test
its strong-field dynamics associated with neutron stars (NSs) and
black holes (BHs)~\cite{ys13}.

Why are we continuously testing GR? First of all, gravity is one of
the most important forces in the Nature whose sophisticated
foundations call for persistent examinations to exquisite
precision. Secondly, puzzles associated with gravity still exist both
theoretically and observationally. From the theoretical viewpoint, GR
fails to make firm physical predictions at the singularities of BHs,
which may need an incorporation of quantum fluctuations to evade
infinities. In a broader concept, there exist fundamental difficulties
in combining GR with quantum principles and quantum field theories of
particle physics into a single unified theory, namely quantum gravity,
due to the issues associated with
nonrenormalizability. Observationally, the phenomena of dark matter
and dark energy could be alluring signals suggesting the breakdown of
GR at galactic and cosmic scales. Several modified alternative gravity
theories beyond GR were proposed to explain these new phenomena as
gravitational manifestations, whose predictions need to be verified or
falsified with further experiments~\cite{cl11,cfps12}. Thirdly, with
persistent tests, high confidence in GR accumulated from observational
facts will reinforce our faith in applying GR under different
circumstances, like to the theories of the Global Positioning
System~\cite{ash03} and the BH accretion disks~\cite{af13}.

In this work, we consider the possibility of local Lorentz invariance
(LLI) violation in the gravity sector. Such a scenario arises numerous
interests recently in the gravity community, for examples, in the
theories of Ho{\v r}ava-Lifshitz gravity~\cite{hor09} and
Einstein-\AE{}ther gravity~\cite{jm01}. In the generic pure-gravity
sector of the standard-model extension (SME) in presence of a
preferred frame (PF), we derive a new limit on the time-time component
of the $\bar s^{\mu\nu}$ matrix in the standard coordinate frame. This
component hardly plays an role in gravity experiments hence it is
difficult to be constrained~\cite{bk06}. Only until recently, Bailey
{\it et al.}  obtained the first empirical limit of $|\bar s^{\rm TT}|
< 3.8\times10^{-3}$ (68\% CL) from the GPB experiment~\cite{beo13}.
The $\bar s^{\rm TT}$ coefficient has no effect at leading order on
the orbital dynamics of binary pulsars if we ignore the relative
velocity of the binary to the PF~\cite{bk06}.  Following the
suggestion in Ref.~\cite{shao14}, we utilize the boost of the pulsar
system with respect to the PF to mix $\bar s^{\rm TT}$ with other
time-spatial and spatial-spatial components of $\bar s^{\mu\nu}$
through a Lorentz transformation. Because binary orbital dynamics is
sensitive to the latter, such a mixture makes a new constraint of
$\bar s^{\rm TT}$ possible. We use the state-of-the-art double-line
observations of neutron star -- white dwarf (NS-WD) binaries with
optical spectroscopy of the former and radio timing of the latter, and
obtain the most stringent limit of $\bar s^{\rm TT}$ up to now. Our
result, $|\bar s^{\rm TT}| < 1.6\times10^{-5}$ (95\% CL), surpasses
the previous limit obtained from GPB by a factor of 500, and further
confirms the validity of GR in its precision in describing
gravitation.

The paper is organized as follows. In the next section, the gravity
sector of SME and its observable effects on the orbital dynamics of
binary pulsars are reviewed. Then the full coordinate transformation
between the Solar system and the binary is elaborated in
section~\ref{sec:coordinate} that is afterwards utilized to mix
different components of $\bar s^{\mu\nu}$. Principles in choosing
suitable binary pulsars for the test are stated in
section~\ref{sec:psr}, and numerical simulations of our three NS-WD
binary systems, namely PSRs~J1738+0333, J1012+5307, and J0348+0432,
are presented in section~\ref{sec:sim}. Discussions on different PFs
and strong-field effects associated with NSs are presented in
section~\ref{sec:discussion}. Section~\ref{sec:summary} briefly
summarizes the work.  The light speed $c=1$ is adopted throughout the
paper.

%=====================================================================
\section{Orbital dynamics of binary pulsars}\label{sec:orbit}
%=====================================================================

The concept of Lorentz symmetry violation is largely motivated by the
hope to probe possible ``relic effects'' at low energy scales from the
new physics of quantum gravity, as well as by the needs to perform the
strictest tests on the most cherished fundamental
principles~\cite{ks89a,ks89b,ck97,ck98,nac69,cn83,kli00}.  With the
fact that GR and the standard model of particle physics have passed
all exquisite empirical examinations up to now~\cite{kr11,will14}, one
would expect that only tiny Lorentz violations are allowed at current
experimental energy scales. Therefore, an effective field theory that
catalogues all possible angles to deviate from an exact Lorentz
symmetry is very helpful in systematically conducting theoretical and
experimental studies. SME is the effort along this direction by
extending our currently well adopted field theories with
Lorentz-violating terms. It initially focused on the matter
sector~\cite{ck97,ck98}, and lately is extended to include the gravity
sector~\cite{kos04,bk06,bkx14}, as well as the couplings between the
matter sector and the gravity sector~\cite{kos04,kt11}.

In SME, a general Lagrangian in Riemann-Cartan spacetime has the
structure ${\cal L} = {\cal L}_{\rm LI} + {\cal L}_{\rm LV}$, where
${\cal L}_{\rm LI}$ and ${\cal L}_{\rm LV}$ are Lorentz-invariant and
Lorentz-violating terms respectively~\cite{kos04}. We here focus o the
limit of Riemannian spacetime and the pure-gravity sector with
Lorentz-violating operators of only mass dimension four or less
(dubbed as the minimal SME, or mSME). With above restrictions, we
have~\cite{bk06},
%---------------------------------------------------------------------
\begin{eqnarray}
  {\cal L}_{\rm LI} &=& \frac{\sqrt{-g}}{16\pi G} (R-2\Lambda) \,, \\
  {\cal L}_{\rm LV} &=& \frac{\sqrt{-g}}{16\pi G} \left(
  -uR + s^{\mu\nu} R^{\rm T}_{\mu\nu} + t^{\kappa\lambda\mu\nu} C_{\kappa\lambda\mu\nu}
  \right) \,,
\end{eqnarray}
%---------------------------------------------------------------------
where $g$ is the determinant of the metric, $R$ is the Ricci scalar,
$R_{\mu\nu}^{\rm T}$ is the trace-free Ricci tensor,
$C_{\kappa\lambda\mu\nu}$ is the Weyl conformal tensor, and $\Lambda$
is the cosmological constant that is set to zero for localized
systems.  

The extra fields, $u$, $s^{\mu\nu}$, and $t^{\kappa\lambda\mu\nu}$,
are dynamical fields that gain vacuum expectation values, $\bar u$,
$\bar s^{\mu\nu}$, and $\bar t^{\kappa\lambda\mu\nu}$, through the
spontaneous symmetry breaking mechanism~\cite{kos04}, which is similar
to the Higgs mechanism in the standard model of particle physics with
a vacuum-condensed scalar field~\cite{hig14,eng14}. In the Riemannian
spacetime with post-Newtonian approximations, consistent treatments
were carried out for the fluctuations around these vacuum expectation
values, including the massless Nambu-Goldstone modes~\cite{bk06}. The
fossilized field $\bar u$ can be absorbed into redefinitions of the
gravitational constant and other fields~\cite{bk06}. We will assume
that proper rescalings are already done hereafter.  The tensor fields,
$\bar s^{\mu\nu}$ and $\bar t^{\kappa\lambda\mu\nu}$, inherit the
symmetries of $R^{\rm T}_{\mu\nu}$ and $C_{\kappa\lambda\mu\nu}$
respectively. It was found that $\bar t^{\kappa\lambda\mu\nu}$ has no
effects on physical experiments at leading order under the simplifying
yet reasonable assumptions made by Bailey and Kosteleck{\'
  y}~\cite{bk06}. Therefore, we will focus on observational effects
from the rescaled vacuum expectation values of $\bar s^{\mu\nu}$ on
the orbital dynamics of binary pulsars. The $\bar s^{\mu\nu}$ field is
traceless and symmetric, consequently, in total nine physical degrees
of freedom are encoded therein.

%---------------------------------------------------------------------
\begin{figure}
  \centering
  \includegraphics[width=9cm]{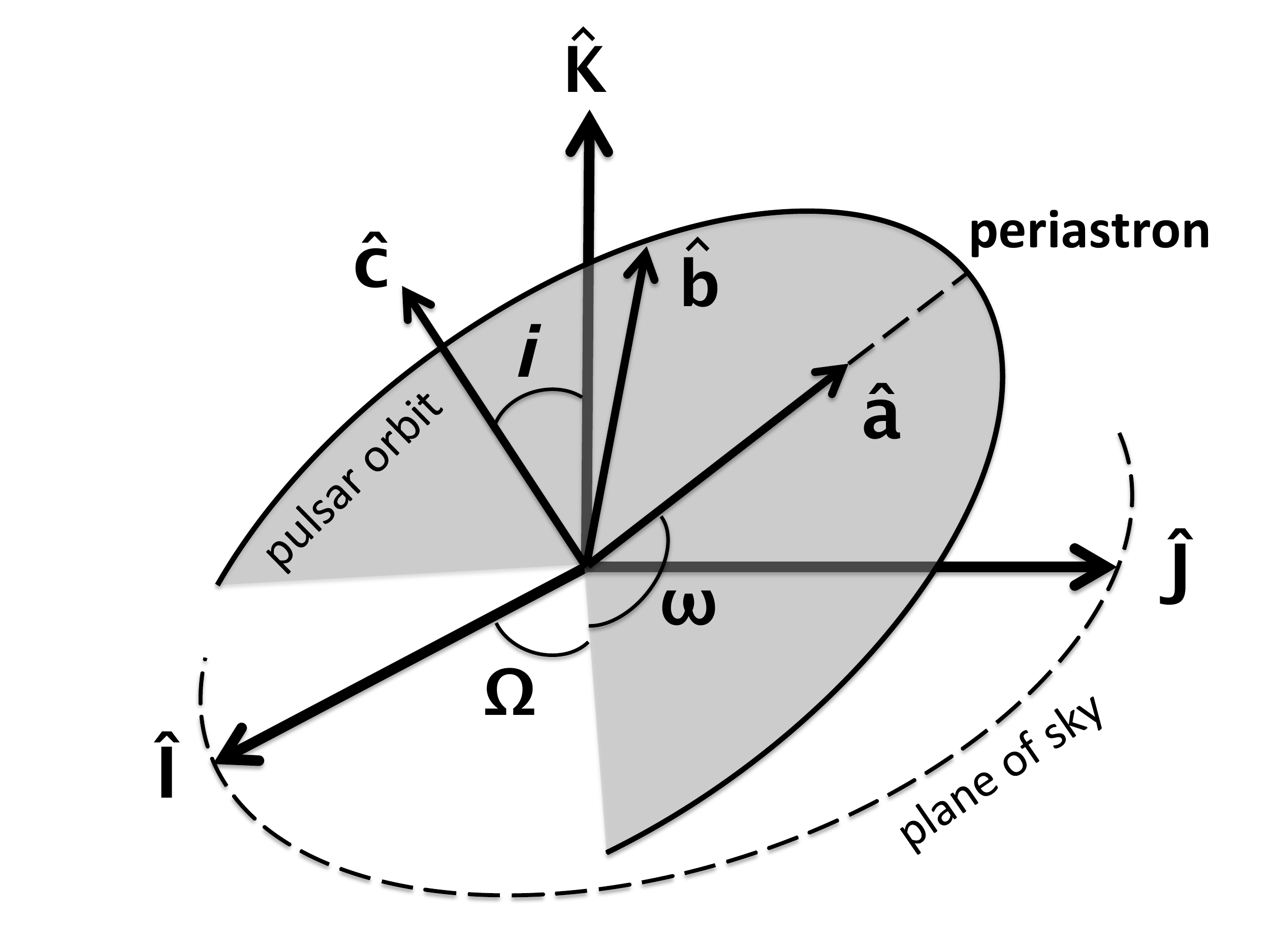}
  \caption{The frame $(\hat{\bf I}, \hat{\bf J}, \hat{\bf K})$ is
    comoving with the pulsar system, with $\hat{\bf K}$ pointing along
    the line of sight to the pulsar from the Earth, while $(\hat{\bf
      I}, \hat{\bf J})$ constitutes the sky plane with $\hat{\bf I}$
    to east and $\hat{\bf J}$ to north. The spatial frame $(\hat{\bf
      a}, \hat{\bf b}, \hat{\bf c})$ is centered at the pulsar system
    with $\hat{\bf a}$ pointing from the center of mass to the
    periastron, $\hat{\bf c}$ along the orbital angular momentum, and
    $\hat{\bf b} \equiv \hat{\bf c} \times \hat{\bf a}$. The frames,
    $(\hat{\bf I}, \hat{\bf J}, \hat{\bf K})$ and $(\hat{\bf a},
    \hat{\bf b}, \hat{\bf c})$, are related through rotation matrices,
    ${\cal R}^{(\Omega)}$, ${\cal R}^{(i)}$, and ${\cal
      R}^{(\omega)}$.\label{fig:geometry}}
\end{figure}
%---------------------------------------------------------------------

In Newtonian gravity, a bound orbit of a binary is a Keplerian
ellipse. For a binary pulsar, the shape of the pulsar orbit is
specified by the semimajor axis, $a$, and the eccentricity, $e$. The
orientation of the orbit with respect to observers is described by
three Euler angles, namely the orbital inclination angle, $i$, the
longitude of periastron, $\omega$, and the longitude of ascending
node, $\Omega$ (see Figure~\ref{fig:geometry} for a graphical
illustration). Kepler's third law relates the orbital size with the
orbital period, $P_b$, through
%---------------------------------------------------------------------
\begin{equation}\label{eq:kepler3}
  G M \propto \left(\frac{2\pi}{P_b}\right)^2 a^3 \,,
\end{equation}
%---------------------------------------------------------------------
where $M \equiv m_1 + m_2$ is the total mass of the system (here we
identify $m_1$ as the pulsar mass and $m_2$ as the companion
mass).\footnote{Since $a$ is defined as the semimajor axis of the {\it
    pulsar} orbit, it should be replaced by $(1+m_1/m_2)a$ to achieve
  an equality in Eq.~(\ref{eq:kepler3}).}  We further define the
orbital frequency, $n_b \equiv 2\pi/P_b$, the projected semimajor axis
of the pulsar orbit, $x \equiv a\sin i$, and the ``characteristic''
orbital velocity, ${\cal V}_{\rm O} \equiv (GMn_b)^{1/3}$ for later
notation simplification.

In presence of Lorentz-violating fossil fields, the orbital dynamics
of a binary pulsar is modified~\cite{bk06,shao14}. By using the
technique of osculating elements, Bailey and Kosteleck{\' y} obtained
the secular changes for orbital elements after averaging over one
orbital period~\cite{bk06},
%---------------------------------------------------------------------
\begin{eqnarray}
  \left\langle \frac{{\rm d} e}{{\rm d} t}\right\rangle
  &=& n_b F_e \sqrt{1-e^2} \left( -eF_e \bar s^{ab} + 2\delta X \,
    {\cal V}_{\rm O} \bar s^{0a}\right) \label{eq:edot} \,, \\
  %------------------------------------------------------------
  \left\langle \frac{{\rm d} \omega}{{\rm d} t}\right\rangle &=&
  \frac{3n_b{\cal V}_{\rm O}^2}{1-e^2} - \frac{n_b F_e \cot
    i}{\sqrt{1-e^2}} \times \\ 
  && \hspace{-1cm} \left(\sin\omega \, \bar s^{ac} +
    \sqrt{1-e^2}\cos\omega \,\bar 
    s^{bc}  + 2\delta X \, e{\cal V}_{\rm O} \cos\omega \, \bar
    s^{0c} \right) \nonumber \\
  && \hspace{-1cm} + n_b F_e\left( F_e \frac{\bar
    s^{aa}-\bar s^{bb}}{2} + \frac{2}{e}\delta X \,  {\cal
      V}_{\rm O}\bar s^{0b} \right) \label{eq:omdot} \,, \nonumber \\
  %------------------------------------------------------------
  \label{eq:xdot} \left\langle \frac{{\rm d} x}{{\rm d}
    t}\right\rangle &=& \frac{1-\delta X}{2}\frac{F_e{\cal
    V}_{\rm O}\cos i}{\sqrt{1-e^2}}  \times  
    \\ &&  \hspace{-1cm} \left(
     \cos\omega \, \bar s^{ac}
    - \sqrt{1-e^2}\sin\omega \, \bar s^{bc}  -2\delta X \,e {\cal
      V}_{\rm O}\sin\omega \, \bar s^{0c} \right) 
    \,,\nonumber 
\end{eqnarray}
%---------------------------------------------------------------------
where we have defined,
%---------------------------------------------------------------------
\begin{eqnarray}
  F_e &\equiv& \frac{1}{1+\sqrt{1-e^2}} \,, \\
  \delta X &\equiv& \frac{m_1-m_2}{m_1+m_2} \,.
\end{eqnarray}
%---------------------------------------------------------------------
The $\bar s^{\mu\nu}$ fields in Eqs.~(\ref{eq:edot}--\ref{eq:xdot})
are defined in the frame that is comoving with the center of the
binary pulsar. The components of $\bar s^{\mu\nu}$ are projected onto
the spatial coordinate frame $(\hat{\bf a}, \hat{\bf b}, \hat{\bf c})$
that is attached with the orbit (see Figure~\ref{fig:geometry}). The
transformation between the pulsar frame and the canonical Sun-centered
celestial-equatorial frame is discussed in
section~\ref{sec:coordinate}.

We will deal with small-eccentricity NS-WD binaries, where in most
cases, the orbits are almost perfectly circular due to mutual tide
forces, frictional dissipation, and exchange of materials during the
evolutionary history~\cite{lk04}. Therefore, instead of $e$ and
$\omega$, two Laplace-Lagrange parameters, $\eta \equiv e\sin\omega$
and $\kappa\equiv e\cos\omega$, are widely used in practice in order
to break the notorious parameter degeneracies~\cite{lcw+01}. In the
limit of $e\ll 1$, it is easy to obtain~\cite{shao14},
%---------------------------------------------------------------------
\begin{eqnarray}
  \left\langle \frac{{\rm d} e}{{\rm d} t}\right\rangle &\simeq& n_b
  \delta X \, {\cal V}_{\rm O} \bar s^{0a} \,,\label{eq:ell1edot} \\
  %-------------------------------------------------------------------
  \left\langle \frac{{\rm d} \omega}{{\rm d} t}\right\rangle &\simeq&
  3n_b{\cal V}_{\rm O}^2 + \frac{n_b}{e} \delta X \, {\cal V}_{\rm O} \bar
  s^{0b} \,, \label{eq:ell1omdot}\\ 
  %-------------------------------------------------------------------
  \label{eq:ell1xdot} \left\langle \frac{{\rm d} x}{{\rm d}
    t}\right\rangle &\simeq& 
  \frac{1-\delta X}{4} {\cal V}_{\rm O}\cos i \left( \bar s^{ac}
  \cos\omega - \bar s^{bc} \sin\omega \right) \,, 
\end{eqnarray}
%---------------------------------------------------------------------
and
%---------------------------------------------------------------------
\begin{eqnarray}
  \left\langle \frac{{\rm d} \eta}{{\rm d}
    t}\right\rangle &\simeq& n_b \delta X {\cal V}_{\rm O} \left( \bar
  s^{0a}\sin\omega + \bar s^{0b}\cos\omega\right) \nonumber \\ 
   & &+ 3e n_b {\cal V}_{\rm O}^2 \cos\omega\,, \label{eq:etadot} \\
  %------------------------------------------------------------
  \left\langle \frac{{\rm d} \kappa}{{\rm d}
    t}\right\rangle &\simeq& n_b \delta X {\cal V}_{\rm O} \left( \bar
  s^{0a}\cos\omega - \bar s^{0b}\sin\omega\right) \nonumber \\
   & &-3en_b {\cal V}_{\rm
    O}^2 \sin\omega\,. \label{eq:kappadot}
\end{eqnarray}
%---------------------------------------------------------------------
These formulae will be used in section~\ref{sec:result} to construct
corresponding tests of gravity.

%=====================================================================
\section{Coordinate transformation}\label{sec:coordinate}
%=====================================================================

In SME, in order to be compatible with the Riemann-Cartan spacetime,
the Lorentz symmetry breaking is spontaneous with underlying dynamical
fluctuations~\cite{kos04}. With this mechanism, the tensorial
background, $\bar s^{\mu\nu}$, is {\it observer} Lorentz-invariant,
while {\it particle} Lorentz-violating. Therefore, to experimentally
probe the magnitudes of $\bar s^{\mu\nu}$ components, an explicit
observer coordinate system should be specified. We adopt the standard
Sun-centered celestial-equatorial frame, $(\hat{\bf T}, \hat{\bf X},
\hat{\bf Y}, \hat{\bf Z})$, in the experimental studies of
SME~\cite{bk06}. In the context of post-Newtonian gravity, it is
chosen as an asymptotically inertial frame that is comoving with the
rest frame of the Solar system and that coincides with the canonical
Sun-centered frame. The $\hat{\bf X}$ axis is pointing from the Earth
to the Sun at vernal equinox of J2000.0 epoch, and the $\hat{\bf Z}$
axis is along the rotating axis of the Earth, and $\hat{\bf Y} \equiv
\hat{\bf Z} \times \hat{\bf X}$ completes a right-handed coordinate
system.

For the purpose of this paper, we will assume the existence of a PF
for simplicity. The PF can be singled out by the global cosmological
evolution or the Universal matter
distribution~\cite{wn72,nw72,will14}. We will keep the choice of PF
general. However, the isotropic CMB frame, which is the most natural
choice from the cosmic viewpoint, is kept in mind as a benchmark. This
simplification, compared with the most generic case of possible
anisotropy in all frames~\cite{bk06,shao14}, is discussed in
section~\ref{sec:discussion}.

In the PF, by virtue of rotational invariance, the $s^{\mu\nu}$ tensor
takes a simple isotropic form~\cite{bk06},
%---------------------------------------------------------------------
\begin{equation}\label{eq:smunu:pf}
  \bar s^{\mu\nu}_{\rm PF} = \bar s^{00}_{\rm PF} \left(
  \begin{array}{cccc}
    1 & 0 & 0 & 0 \\
    0 & \frac{1}{3} & 0 & 0 \\
    0 & 0 & \frac{1}{3} & 0 \\
    0 & 0 & 0 & \frac{1}{3} \\
  \end{array}
  \right) \,.
\end{equation}
%---------------------------------------------------------------------
Here for numerical reasons, we have denoted $\bar s^{00}_{\rm PF}$ to
be the $\bar s^{00}$ component in the PF. Although the $\bar s^{00}$
component of $\bar s^{\mu\nu}$ will in general change under a
coordinate transformation, the value of $\bar s^{00}_{\rm PF}$, which
specified in the PF, is fixed. Worthy to mention that, we also
naturally have $\bar t^{\kappa\lambda\mu\nu} = 0$ in the
PF~\cite{bk06}.

If we consider a frame that is moving with respect to the PF with a
velocity ${\bf w} \equiv (\wx,\wy,\wz)$, the $s^{\mu\nu}$ matrix takes
the form (see Eq.~(68) in Ref.~\cite{bk06}),
%---------------------------------------------------------------------
\begin{equation}\label{eq:boost}\label{eq:smunu}
  \bar s^{\mu\nu} = \bar s^{\mu\nu}_{\rm PF} + \bar s^{\mu\nu}_{\bf w}  \,,
\end{equation}
%---------------------------------------------------------------------
where 
%---------------------------------------------------------------------
\begin{equation}\label{eq:sw}
  \bar s^{\mu\nu}_{\bf w} = \frac{4}{3} \bar s^{00}_{\rm PF} \left(
  \begin{array}{cccc}
    \wx^2+\wy^2+\wz^2 & -\wx & -\wy & -\wz \\
    -\wx & \wx^2 & \wx\wy & \wx\wz \\
    -\wy & \wx\wy & \wy^2 & \wy\wz \\
    -\wz & \wx\wz & \wy\wz & \wz^2 \\
  \end{array}
  \right) \,.
\end{equation}
%---------------------------------------------------------------------

Previous tests of the pure-gravity sector of SME with pulsar
observations, no matter with the orbital dynamics of binary
pulsars~\cite{bk06,sw12,shao14} or the spin evolution of solitary
pulsars~\cite{sck+13,shao14}, include no observable effect from the
component $\bar s^{00} \equiv \bar s^{11} + \bar s^{22} + \bar
s^{33}$, under the assumption that the Lorentz boost in
Eq.~(\ref{eq:boost}) is negligible~\cite{bk06}. Therefore, we were
only able to constrain the other eight time-spatial and
spatial-spatial components of $\bar s^{\mu\nu}$, even with
over-abounded twenty-seven independent tests in
Ref.~\cite{shao14}. Nevertheless, with the Lorentz boost, one can
clearly see a mixture between the $\bar s^{00}$ component and the
$\bar s^{0j}$ and $\bar s^{jk}$ components ($j,k=1,2,3$). Although the
systematic velocity of a binary pulsar is small (typically, $|{\bf w}|
\sim {\cal O}(10^{-3})$), precision experiments with pulsars can still
put a meaningful constraint on the $\bar s^{00}$ component with
careful studies~\cite{shao14}. This is the main idea of this work that
establishes the basis of the test below.

It is worthy to mention that, in the standard post-Newtonian frame of
SME, we are interested in constraining the component $\bar s^{\rm
  TT}$, which is the $00$-component of $\bar s^{\mu\nu}$ in the
standard Sun-centered celestial-equatorial frame, that is
%---------------------------------------------------------------------
\begin{equation}
  \bar s^{\rm TT} = \bar s^{00}_{\rm PF} \left( 1 + 
  \frac{4}{3} {\bf w}_\odot^2 \right) \,,
\end{equation}
%---------------------------------------------------------------------
where ${\bf w}_\odot$ is the velocity of the Solar system with respect
to the PF. This rescaling is negligible, nevertheless, it is accounted
for in our calculations.

Besides the boost in Eq.~(\ref{eq:boost}), one also needs to perform a
spatial rotation, ${\cal R}$, to align the spatial axes of $(\hat{\bf
  a},\hat{\bf b},\hat{\bf c})$ and $(\hat{\bf X},\hat{\bf Y},\hat{\bf
  Z})$,
%---------------------------------------------------------------------
\begin{equation}
  \left(
  \begin{array}{c}
    \hat{\bf a} \\
    \hat{\bf b} \\
    \hat{\bf c}
  \end{array}
  \right) = 
  {\cal R}
  \left(
  \begin{array}{c}
    \hat{\bf X} \\
    \hat{\bf Y} \\
    \hat{\bf Z}
  \end{array}
  \right) \,.
\end{equation}
%---------------------------------------------------------------------
With the help of an intermediate coordinate system $(\hat{\bf
  I},\hat{\bf J},\hat{\bf K})$ in Figure~\ref{fig:geometry}, one can
decompose the full rotation into five simple parts~\cite{shao14},
%---------------------------------------------------------------------
\begin{equation}
  {\cal R} = {\cal R}^{(\omega)} {\cal R}^{(i)} {\cal R}^{(\Omega)}
  {\cal R}^{(\delta)} {\cal R}^{(\alpha)} \,,
\end{equation}
%---------------------------------------------------------------------
with
%---------------------------------------------------------------------
\begin{eqnarray}
  {\cal R}^{(\alpha)} &=& 
  \left(
  \begin{array}{ccc}
    -\sin\alpha & \cos\alpha & 0 \\
    -\cos\alpha & -\sin\alpha & 0 \\
    0 & 0 & 1
  \end{array}
  \right) \,,\\
  %-----------------------------------------------------------
  {\cal R}^{(\delta)} &=& 
    \left(
  \begin{array}{ccc}
    1 & 0 & 0 \\
    0 & \sin\delta & \cos\delta \\
    0 & -\cos\delta & \sin\delta
  \end{array}
  \right) \,,\\
  %-----------------------------------------------------------
  {\cal R}^{(\Omega)} &=&
    \left(
  \begin{array}{ccc}
    \cos\Omega & \sin\Omega & 0 \\
    -\sin\Omega & \cos\Omega & 0 \\
    0 & 0 & 1
  \end{array}
  \right) \,,\\
  %-----------------------------------------------------------
  {\cal R}^{(i)} &=& 
    \left(
  \begin{array}{ccc}
    1 & 0 & 0 \\
    0 & \cos i & \sin i \\
    0 & -\sin i & \cos i
  \end{array}
  \right) \,,\\
  %-----------------------------------------------------------
  {\cal R}^{(\omega)} &=& 
    \left(
  \begin{array}{ccc}
    \cos\omega & \sin\omega & 0 \\
   -\sin\omega & \cos\omega & 0 \\
    0 & 0 & 1
  \end{array}
  \right) \,,
\end{eqnarray}
%---------------------------------------------------------------------
where $\alpha$ and $\delta$ are the right ascension and declination of
the binary pulsar.

Instead of performing such a rotation to $\bar s^{\mu\nu}$ in
Eq.~(\ref{eq:smunu}) or $\bar s^{\mu\nu}_{\bf w}$ in
Eq.~(\ref{eq:sw})\footnote{Apparently, $\bar s^{\mu\nu}_{\rm PF}$ in
  Eq.~(\ref{eq:smunu:pf}) does not change under a spatial rotation
  ${\cal R}$.}, one can decompose the velocity ${\bf w}$ in the
$(\hat{\bf a},\hat{\bf b},\hat{\bf c})$ coordinate frame,
%---------------------------------------------------------------------
\begin{eqnarray}
  {\bf w} &=& \wx \hat{\bf X} + \wy \hat{\bf Y} + \wz \hat{\bf Z}
  \nonumber \\
  &=& w_a \hat{\bf a} + w_b \hat{\bf b} + w_c \hat{\bf c} \,,
\end{eqnarray}
%---------------------------------------------------------------------
and replace $(\wx,\wy,\wz)$ with $(w_a,w_b,w_c)$ in Eq.~(\ref{eq:sw})
to obtain the desired form of $\bar s^{\mu\nu}$ in the comoving frame
of the pulsar system with the spatial axes $(\hat{\bf a},\hat{\bf
  b},\hat{\bf c})$.  To be more explicit, the components of the $\bar
s^{\mu\nu}$ field that are to be used in
Eqs.~(\ref{eq:edot}--\ref{eq:kappadot}) are~\cite{bk06},
%---------------------------------------------------------------------
\begin{eqnarray}
  \bar s^{0a} &=& -\frac{4}{3} \bar s^{00}_{\rm PF} w_a
  \,, \label{eq:s0a} \\ 
  \bar s^{0b} &=& -\frac{4}{3} \bar s^{00}_{\rm PF} w_b \,, \\
  \bar s^{0c} &=& -\frac{4}{3} \bar s^{00}_{\rm PF} w_c \,, \\
  \bar s^{ab} &=& \frac{4}{3} \bar s^{00}_{\rm PF} w_a w_b \,, \\
  \bar s^{bc} &=& \frac{4}{3} \bar s^{00}_{\rm PF} w_b w_c \,, \\
  \bar s^{ac} &=& \frac{4}{3} \bar s^{00}_{\rm PF} w_a w_c \,, \\
  \bar s^{aa} &=& \frac{1}{3} \bar s^{00}_{\rm PF} (1+4w_a^2) \,, \\
  \bar s^{bb} &=& \frac{1}{3} \bar s^{00}_{\rm PF} (1+4w_b^2) \,, \\
  \bar s^{cc} &=& \frac{1}{3} \bar s^{00}_{\rm PF} (1+4w_c^2)
  \,, \label{eq:scc} 
\end{eqnarray}
%---------------------------------------------------------------------
where, to reiterate, $(w_a,w_b,w_c)$ are the components of the
3-dimensional velocity of the pulsar system with respect to the PF,
projected on the coordinate frame $(\hat{\bf a},\hat{\bf b},\hat{\bf
  c})$.

%=====================================================================
\section{Simulations and Results}\label{sec:result}
%=====================================================================

%---------------------------------------------------------------------
\begin{table*}
\caption{Relevant quantities of PSRs~J1012+5307 \cite{lwj+09},
  J1738+0333 \cite{fwe+12}, and J0348+0432~\cite{afw+13} for the test,
  from radio timing and optical spectroscopy
  observations. Parenthesized numbers represent the $1$-$\sigma$
  uncertainty in the last digits quoted. The listed Laplace-Lagrange
  parameter, $\eta$, is the {\it intrinsic} value, after subtraction
  of the contribution from the Shapiro delay~\cite{lcw+01}. For
  orbital inclination, there is an ambiguity between $i$ and
  $180^\circ-i$; only the value $i < 90^\circ$ is
  tabulated. \label{tab:psr}}
  \begin{tabular}{p{6.5cm}p{3.5cm}p{3.5cm}p{3.5cm}}
    \hline\hline
     & PSR~J1012+5307 & PSR~J1738+0333 & PSR~J0348+0432 \\
    \hline
    \multicolumn{4}{l}{\tt Observed Quantities} \\
    \hline
    Observational span, $T_{\rm obs}$ (year) & $\sim15$~\cite{lwj+09} &
    $\sim10$~\cite{fwe+12} & $\sim4$~\cite{afw+13} \\
    Right ascension, $\alpha$ (J2000) &
      ${\rm 10^h12^m33^s \hspace{-1.2mm}.  4341010(99)}$ &
      ${\rm 17^h38^m53^s \hspace{-1.2mm}. 9658386(7)}$ &
      ${\rm 03^h48^m43^s \hspace{-1.2mm}. 639000(4)}$ \\
    Declination, $\delta$ (J2000) &
      $53^\circ07^\prime02^{\prime\prime} \hspace{-1.2mm} .  60070(13)$ & 
      $03^\circ33^\prime10^{\prime\prime} \hspace{-1.2mm} .  86667(3)$&
      $04^\circ32^\prime11^{\prime\prime} \hspace{-1.2mm} .  4580(2)$ \\
    Proper motion in $\alpha$, $\mu_\alpha~(\textrm{mas\,yr}^{-1})$ & 
    2.562(14) & 7.037(5) & 4.04(16) \\
    Proper motion in $\delta$, $\mu_\delta~(\textrm{mas\,yr}^{-1})$ & 
    $-$25.61(2) & 5.073(12) & 3.5(6) \\
    Distance, $d$ (kpc) & $0.836(80)$ & $1.47(10)$ & $2.1(2)$ \\
    Radial velocity, $v_r$ (km\,s$^{-1}$) & $44(8)$ & $-42(16)$ & $-1(20)$ \\
    Spin period, $P$ (ms) & 5.255749014115410(15) &
    5.850095859775683(5) & 39.1226569017806(5) \\
    Orbital period, $P_{\rm b}$ (day) & 0.60467271355(3) & 0.3547907398724(13)
    & 0.102424062722(7) \\
    Projected semimajor axis, $x$ (lt-s) & 0.5818172(2) &
    0.343429130(17) & 0.14097938(7) \\
    $\eta \equiv e \sin\omega~(10^{-7})$ & $-1.4(34)$ &  
    $-1.4(11)$ & $19(10)$ \\
    $\kappa \equiv e \cos\omega~(10^{-7})$ & $0.6(31)$ & $3.1(11)$
    & $14(10)$ \\
    Time derivative of $x$, $\dot{x}~(10^{-15}~\textrm{s\,s}^{-1})$ & 2.3(8) & 
    0.7(5) & $\cdots$ \\
    Mass ratio, $q \equiv m_1/m_2$ & 10.5(5) & 8.1(2) & 11.70(13) \\
    Companion mass, $m_2~({\rm M}_\odot)$ & 0.16(2) &
    $0.181^{+0.008}_{-0.007}$ & 0.172(3) \\
    Pulsar mass, $m_1~(\textrm{M}_\odot)$ & 1.64(22) &
    $1.46^{+0.06}_{-0.05}$ & 2.01(4) \\
    $\delta X \equiv  (q-1)/(q+1)$ & 0.826(8) & 0.780(5) &
    0.843(2) \\
    \hline
    \multicolumn{4}{l}{\tt Estimated Quantities} \\
    \hline
    Upper limit of $|\dot x|~(10^{-15}~\textrm{s\,s}^{-1})$ & $\cdots$
    & $\cdots$ & 1.9 \\
    Upper limit of $|\dot\eta|$ ($10^{-14}\,{\rm s}^{-1}$) & 0.25 & 0.12
    & 2.7 \\
    Upper limit of $|\dot\kappa|$ ($10^{-14}\,{\rm s}^{-1}$) & 0.23 &
    0.12 & 2.7 \\
    \hline
    \multicolumn{4}{l}{\tt Derived Quantities Based on GR} \\
    \hline
    Orbital inclination, $i$ (deg) &  52(4) & 32.6(10) & 40.2(6) \\
    Advance of periastron, $\dot{\omega}\,({\rm
      deg\,yr}^{-1})$ & 0.69(6) & 1.57(5) & 14.9(2) \\
    Characteristic velocity, $\mathcal{V}_{\rm O}\,({\rm
      km\,s}^{-1})$ & 308(13) & 355(5) & 590(4) \\
    \hline
  \end{tabular}
\end{table*}
%---------------------------------------------------------------------

%---------------------------------------------------------------------
\begin{figure}
  \centering
  \includegraphics[width=9cm]{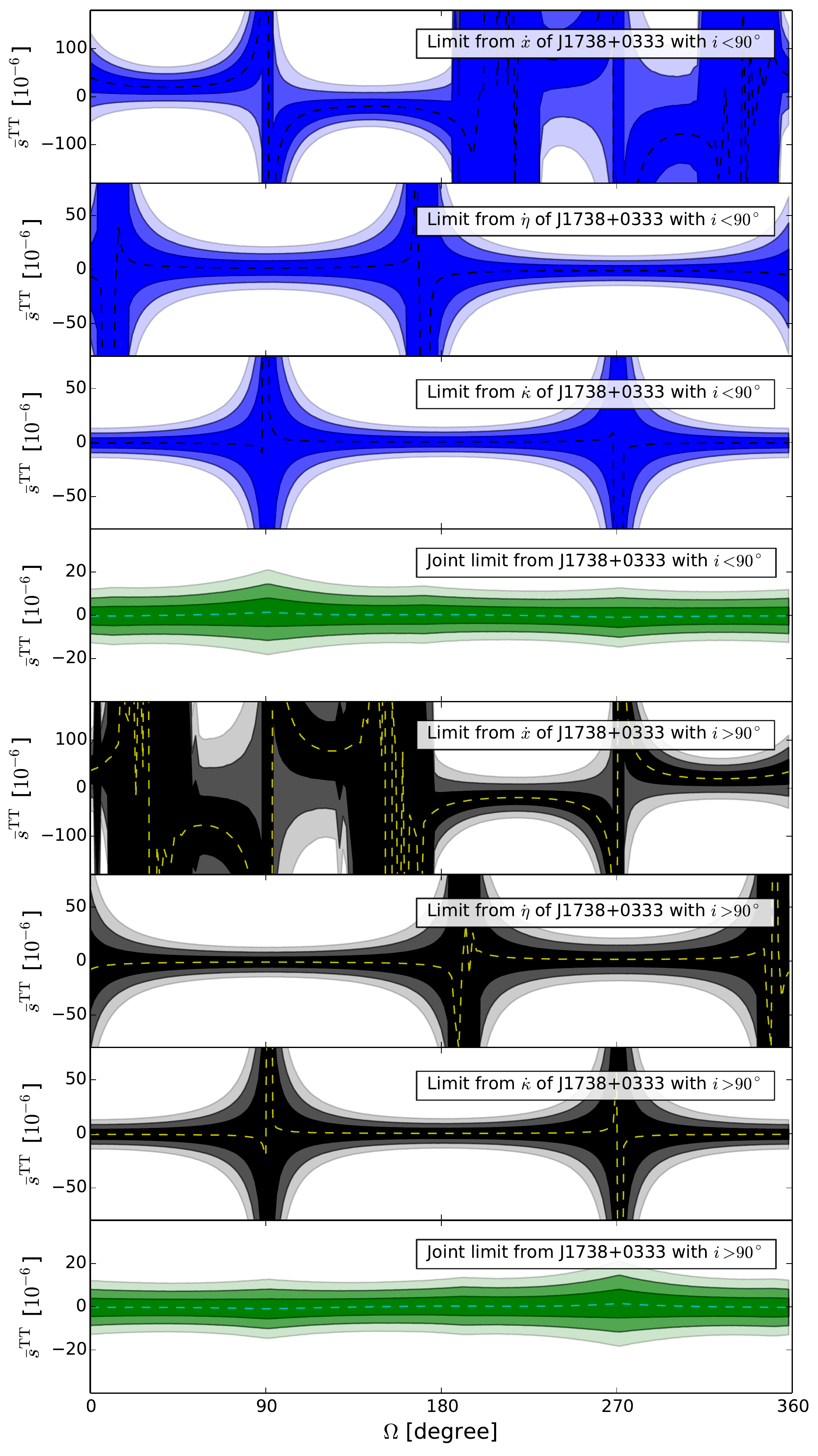}
  \caption{Limits on $\bar s^{\rm TT}$ from PSR~J1738+0333. In each
    panel, 1-$\sigma$, 2-$\sigma$, and 3-$\sigma$ contours are drawn
    with different color scales, while dashed lines are the central
    values from our Monte Carlo simulations. The upper four panels
    show in sequence the limits from $\dot x$, $\dot \eta$, $\dot
    \kappa$ and their combination, with an orbital inclination $i <
    90^\circ$. The cases for $i>90^\circ$ are depicted in the lower
    four panels. \label{fig:J1738}}
\end{figure}
%---------------------------------------------------------------------

%---------------------------------------------------------------------
\subsection{Pulsar systems}\label{sec:psr}
%---------------------------------------------------------------------

%---------------------------------------------------------------------
\begin{figure}
  \centering
  \includegraphics[width=9cm]{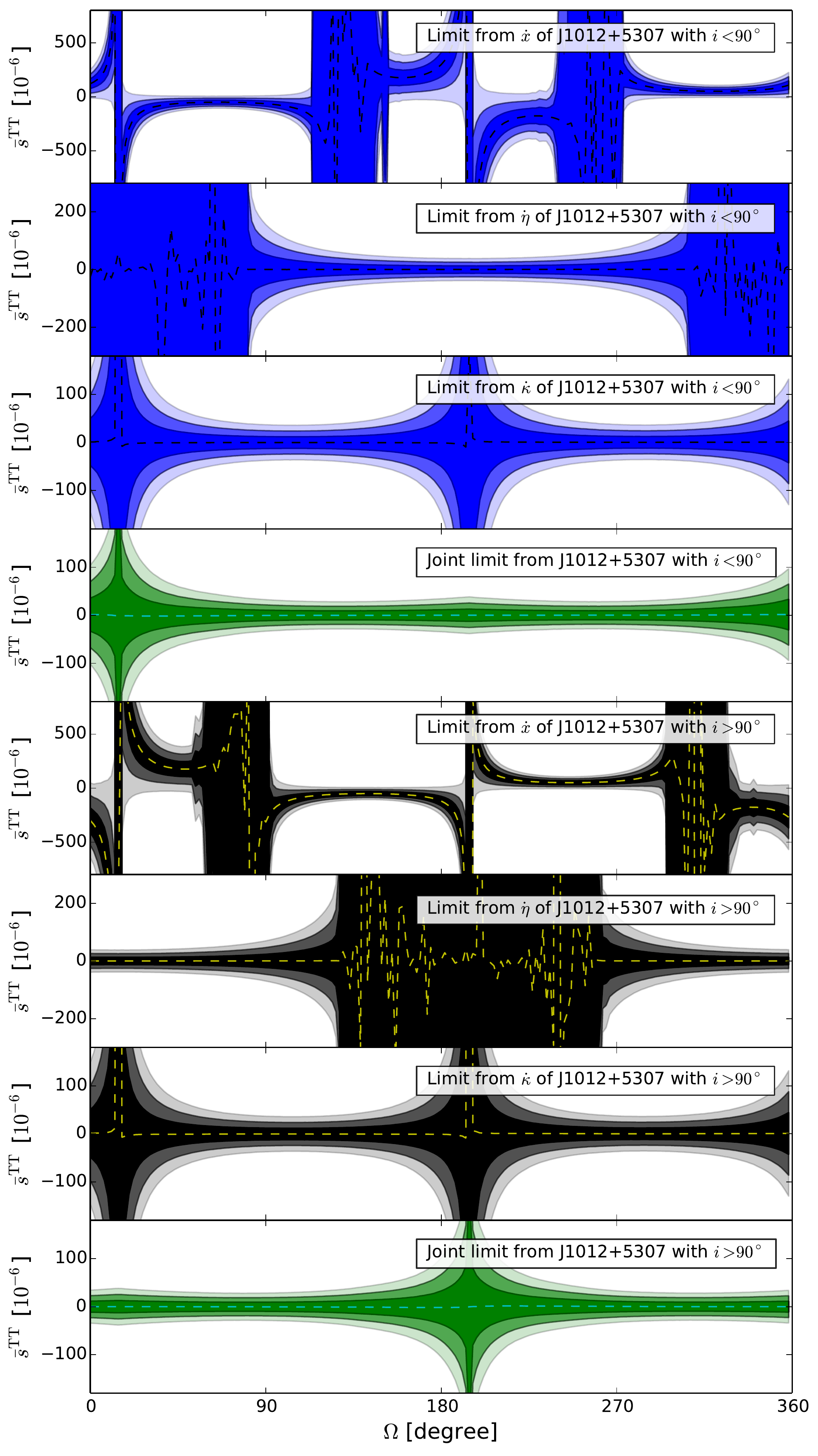}
  \caption{Same as Figure~\ref{fig:J1738}, for
    PSR~J1012+5307. \label{fig:J1012}}
\end{figure}
%---------------------------------------------------------------------

In order to perform tests of the $\bar s^{\rm TT}$ component in the
pure-gravity sector of SME with binary pulsars, there are several
observational requirements that need to be met.
\begin{itemize}
\item First of all, because we need the boost to mix different
  components in the $\bar s^{\mu\nu}$ matrix, a measurement of the
  3-dimensional velocity of the pulsar system is required. Usually,
  for a well-timed binary pulsar that is not too far away from the
  Solar system, we can obtain its proper motion after several years of
  radio timing. Together with the distance information from the
  parallax measurement from radio timing, or alternatively, from Very
  Long Baseline Interferometry (VLBI), we can get the system's
  2-dimensional motion projected on the sky plane. In general, the
  systematic velocity of the binary along its line of sight is not
  measurable in radio timing. Fortunately, for some NS-WD systems, we
  can use the orbitally phase-resolved optical spectroscopy of the
  white dwarf to separate its sinusoidally varying (projected) orbital
  velocity and its nearly constant systematic radial
  velocity~\cite{cgk98,akk+12,afw+13}. We will make use of three
  small-eccentricity NS-WD binaries with radial velocity measurements,
  namely PSRs~J1012+5307~\cite{lwj+09}, J1738+0333~\cite{fwe+12} and
  J0348+0432~\cite{afw+13}.
\item As can be seen from Eqs.~(\ref{eq:edot}--\ref{eq:xdot}), we
  demand the measurements or upper limits of $\dot e$, $\dot\omega$
  and $\dot x$ to perform these tests. Usually, a very good timing
  precision is needed to achieve these observations. Therefore, only
  very well timed binary pulsars are considered here (see
  Table~\ref{tab:psr}). For small-eccentricity binary pulsars, in
  practice, we use the Laplace-Lagrange parameters, $\eta$ and
  $\kappa$, to break parameter degeneracies within the fitting
  procedure of times of arrival of radio pulse
  signals~\cite{lcw+01}. Therefore, for these pulsars, we will use
  $\dot\eta$ and $\dot\kappa$ in
  Eqs.~(\ref{eq:etadot}--\ref{eq:kappadot}) instead of $\dot e$ and
  $\dot\omega$. Because for some pulsars, $\dot x$, $\dot\eta$, and
  $\dot \kappa$ were not reported along with other timing parameters
  in their timing solutions published in literature, we adopt the
  methodology in Ref.~\cite{shao14}. Whenever inaccessible, we
  conservatively estimate 68\% CL upper limits for these parameters as
  $|\dot{\cal P}|^{\rm upper} = \sqrt{12} \sigma_{\cal P} / T_{\rm
    obs}$ (${\cal P} = x, \eta, \kappa$), where $T_{\rm obs}$ is the
  time span used in deriving the timing solution. This choice is in
  accordance with the case of linear-in-time evolution. It is
  justified, because if there is any large effect from Lorentz
  violation or other new sources, the changes in these parameters
  should have been detected already in these systems, or the
  uncertainties of $x$, $\eta$, and $\kappa$ derived from times of
  arrival of pulse signals cannot be too
  minuscule~\cite{shao14}. Nevertheless, to fully account for all
  parameter correlations, one will need refitting of times of arrival
  of these pulsars explicitly with parameters, $\dot x$, $\dot \eta$,
  and $\dot\kappa$, in the timing model.
\item The test of $\bar s^{\rm TT}$ is possible only if component
  masses of the binary are measured independently of the timing
  parameters we are using in the test. Interestingly, such mass
  measurements were already obtained, thanks to the optical
  observations of the WD companions, with the three small-eccentricity
  binary pulsars we are to use~\cite{cgk98,lwj+09,akk+12,afw+13}. From
  optical observations, the mass of the WD can be inferred based on
  the WD atmosphere models and, from the ratio of the (projected)
  orbital velocities of the WD (from optical spectroscopy) and the
  pulsar (from radio timing), the mass ratio of two components can be
  derived. Thereby we can get two component masses without assuming GR
  to be the correct underlying gravity theory as done in other systems
  with post-Keplerian parameters~\cite{sta03,lk04,wex14}. With
  component masses, one can estimate the mass difference, $\delta X$,
  and the characteristic orbital velocity of the system, ${\cal
    V}_{\rm O}$, basing on the Kepler's third law.\footnote{Here in
    Kepler's third law, we use the gravitational constant measured in
    the weak field, namely, the {\it Cavendish $G$}, which is
    justified by other pulsar systems (see {\it e.g.} Ref.~\cite{kw09}).}
\item The geometry of the binary orbit, in terms of three Euler
  angles, $i$, $\omega$ and $\Omega$, is also required to project the
  3-dimensional velocity, ${\bf w}$, onto the coordinate system
  $(\hat{\bf a}, \hat{\bf b}, \hat{\bf c})$. The orbital inclination
  is calculable from Kepler's law since two masses of the system are
  known. Because we can only infer $\sin i$ from radio timing, there
  is an ambiguity between $i$ and $180^\circ - i$.  The longitude of
  periastron, $\omega$, is also calculable once $\eta$ and $\kappa$
  are given.  However, the longitude of ascending node, $\Omega$, is
  generally not an observable. In Ref.~\cite{shao14}, we had to treat
  it as an unknown quantity uniformly distributed in the range
  $[0^\circ,360^\circ )$, that renders the tests as {\it
      probabilistic} tests. In this work, we will develop a {\it
      robust} test for $\bar s^{\rm TT}$ even without knowing the
    actual value of $\Omega$. The idea is similar to the robust test
    of the strong-field parameter, $\hat\alpha_1$, in the parametrized
    post-Newtonian (PPN) formalism, in Ref.~\cite{sw12}, and will be
    elaborated in detail in the next subsection (also see
    Ref.~\cite{fkw12} for a similar idea in testing of strong
    equivalence principle).
\end{itemize}

As mentioned, to meet the requirements listed above, we choose
PSRs~J1012+5307~\cite{lwj+09}, J1738+0333~\cite{fwe+12}, and
J0348+0432~\cite{afw+13} to perform the test of $\bar s^{\rm
  TT}$. They are all well-timed small-eccentricity NS-WD binaries with
both radio and optical observations.  Besides, these binaries are also
relativistic binaries with small orbital periods ($P_b <
15\,$hours). This is important, because relativistic orbits boost the
figures of merit of the test with larger orbital frequency, $n_b$, and
larger characteristic orbital velocity, ${\cal V}_{\rm O}$ (see
Eqs.~(\ref{eq:ell1xdot}--\ref{eq:kappadot})).  Relevant parameters for
the test of these systems, from radio timing and optical spectroscopy,
are tabulated in Table~\ref{tab:psr} (see the original
publications~\cite{fwe+12,lwj+09,afw+13} for details).

%---------------------------------------------------------------------
\subsection{Constraints on $\bar s^{\rm TT}$}\label{sec:sim}
%---------------------------------------------------------------------

%---------------------------------------------------------------------
\begin{figure}
  \centering
  \includegraphics[width=9cm]{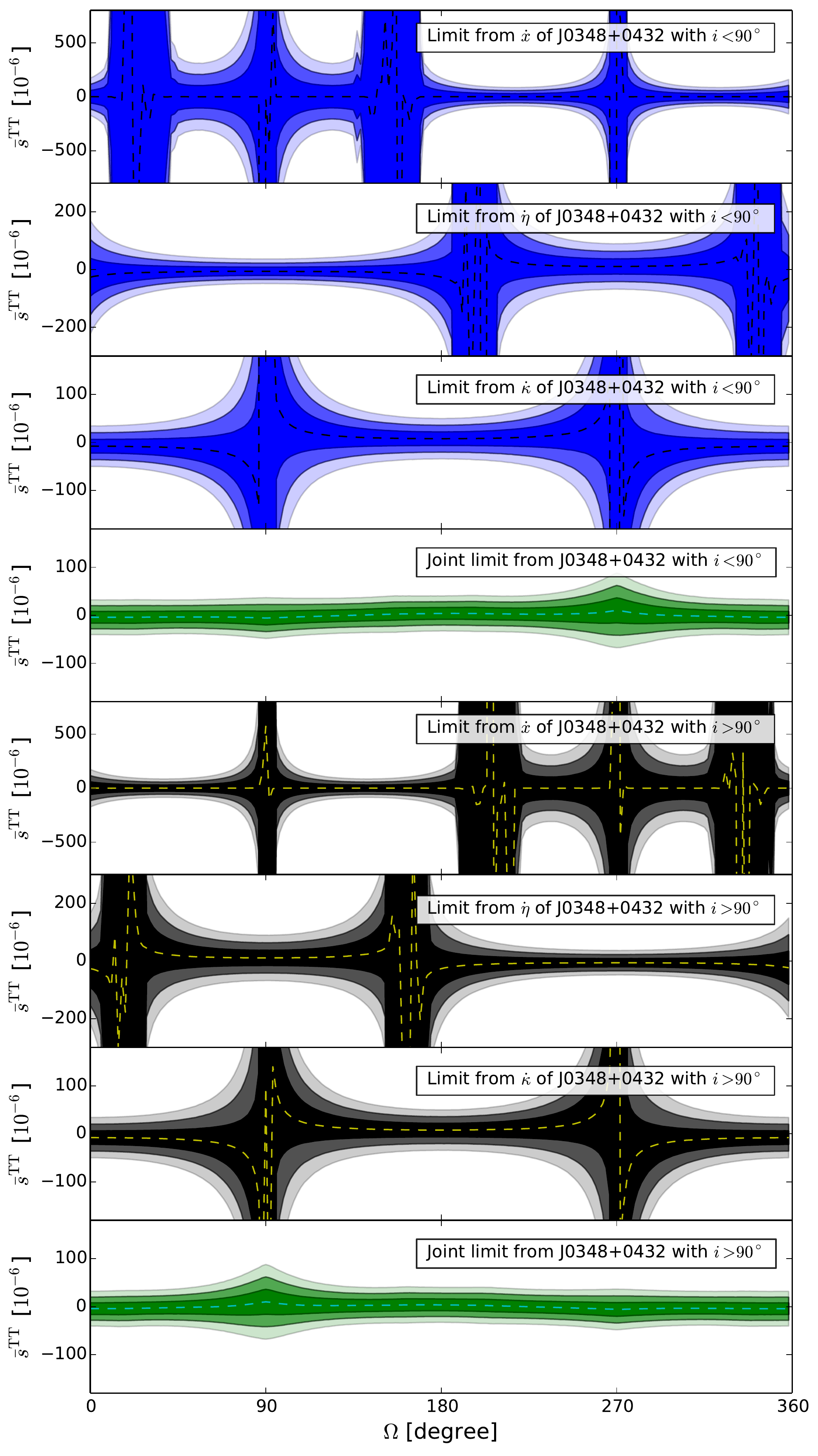}
  \caption{Same as Figure~\ref{fig:J1738}, for
    PSR~J0348+0432. \label{fig:J0348}}
\end{figure}
%---------------------------------------------------------------------

We use expressions of $\dot x$, $\dot\eta$, and $\dot \kappa$ in
Eqs.~(\ref{eq:ell1xdot}--\ref{eq:kappadot}), together with
Eqs.~(\ref{eq:s0a}--\ref{eq:scc}), to perform tests of $\bar s^{\rm
  TT}$. Here we assume that the isotropic CMB frame singles out a PF
(however, see generalized cases in the next section). Therefore, the
``absolute'' velocity of the pulsar system, ${\bf w} = {\bf w}_\odot +
{\bf v}$, is a vectorial superposition of the velocity of the Solar
system to the CMB frame, ${\bf w}_\odot$, and the velocity of the
pulsar system to the Solar system, ${\bf v}$. From Wilkinson Microwave
Anisotropy Probe (WMAP) experiments, a CMB dipole measurement of
$3.355\pm0.008\,$mK was obtained, which implies a peculiar velocity of
the Solar system barycentre, $|{\bf w}_\odot| =
369.0\pm0.9\,$km\,s$^{-1}$, in the direction of Galactic longitude and
latitude, $(l,b) = (263.99^\circ \pm 0.14^\circ,
48.26^\circ\pm0.03^\circ)$~\cite{hlk+13}. The 3-dimensional velocities
of binary systems, ${\bf v}$'s, are derived from joint radio timing
and optical spectroscopy observations. Direct calculations show that
the absolute velocities of binary pulsars are of ${\cal O}(10^{-3})$
for three binary systems in Table~\ref{tab:psr}.

Because the longitude of ascending node, $\Omega$, is unknown for all
three pulsar systems, we scan through its values in the range
$[0^\circ, 360^\circ)$. After choosing the orbital inclination between
  $i$ and $180^\circ-i$, for each given $\Omega$, we can fix the
  absolute orientation of the coordinate frame $(\hat{\bf a}, \hat{\bf
    b}, \hat{\bf c})$ at Newtonian order.  The absolute velocity of
  the binary is projected on the $(\hat{\bf a}, \hat{\bf b}, \hat{\bf
    c})$ frame to get its coordinate components, $(w_a, w_b,
  w_c)$. Having all these information at hand, from the measured or
  reasonably estimated $\dot x$, $\dot \eta$, and $\dot \kappa$ (see
  Table~\ref{tab:psr}), we can calculate limits of $\bar s^{\rm TT}$
  from Eqs.~(\ref{eq:ell1xdot}--\ref{eq:kappadot}) for each $\Omega$.

For each binary pulsar, we perform $10^5$ Monte Carlo simulations for
each value of $\Omega$ to account for the observational uncertainties
in parameters listed in Table~\ref{tab:psr}.  From the results of
these simulations, we can read out the constraints of $\bar s^{\rm
  TT}$ from $\dot x$, $\dot\eta$, and $\dot \kappa$ separately. The
result from PSR~J1738+0333 is illustrated in the first three panels of
Figure~\ref{fig:J1738} for $\dot x$, $\dot\eta$, and $\dot \kappa$, as
a function of $\Omega$ with an orbital inclination $i<90^\circ$
(namely, $i=32.6^\circ \pm 1.0^\circ$). Different colors correspond to
68.3\%, 95.5\%, and 99.7\% CLs. These results remind us the test of
the strong-field PPN parameter, $\hat\alpha_2$, in Ref.~\cite{sw12}
(see their Figures 2--3), where for some $\Omega$'s, the parameter
values are basically unconstrained due to ``unfavored'' geometrical
configurations.  This is due to the vectorial/tensorial nature of the
LLI violation.  It happens for the $\dot x$ test in Ref.~\cite{sw12},
and it is still true for all three tests here (see Ref.~\cite{sw12}
for more discussions).

For the $\hat\alpha_2$ test in Ref.~\cite{sw12} there is only one
``anomalous'' parameter entering the test, namely $\dot x$, while here
we have three parameters entering. Therefore, for a given $\Omega$, if
an $\bar s^{\rm TT}$ is to pass the test, it should pass all three
tests {\it simultaneously}. In other words, for a given $\Omega$, we
can adopt the tightest constraint of $\bar s^{\rm TT}$, out of the
three limits from $\dot x$, $\dot \eta$, and $\dot \kappa$. Such
limits from PSR~J1738+0333 with $i<90^\circ$ is depicted in the fourth
panel of Figure~\ref{fig:J1738}. We can see that, because the unbound
situations do not occur simultaneously for {\it all} three tests for
{\it any} given $\Omega$, a quite smooth limit of $\bar s^{\rm TT}$
can be attained as a function of $\Omega$. Since there exists no
measurement of $\Omega$ for PSR~J1738+0333 yet, we conservatively read
out the worst constraint from the fourth panel of
Figure~\ref{fig:J1738}, that gives,
%---------------------------------------------------------------------
\begin{equation}\label{eq:limit:J1738}
  |\bar s^{\rm TT}| < 1.6 \times 10^{-5}  \quad (95\%~{\rm CL}) \,,
\end{equation}
%---------------------------------------------------------------------
at $\Omega \simeq 90^\circ$.

We also plot the case for $i > 90^\circ$ (namely, $i = 147.4^\circ\pm
1.0^\circ$) in the last four panels of Figure~\ref{fig:J1738} for
PSR~J1738+0333. It is symmetric with respect to the case of $i <
90^\circ$, and the worst limit from the joint constraint (the eighth
panel) also gives $|\bar s^{\rm TT}| < 1.6 \times 10^{-5} ~ (95\%~{\rm
  CL})$, at $\Omega \simeq 270^\circ$. We claim that this limit is
{\it robust} instead of {\it probabilistic}, in contrast to the limits
presented in Ref.~\cite{shao14} where $\Omega$'s were effectively
``averaged'' out in the range $[0^\circ,360^\circ)$. The limit in
  Eq.~(\ref{eq:limit:J1738}) is about 500 times better than the
  current best (yet unique) limit from the GPB
  experiment~\cite{beo13}.

The constraints from PSR~J1012+5307 are illustrated in
Figure~\ref{fig:J1012} as a function of $\Omega$. In this system, it
is seen that for the joint limit, there still exist unconstrained
regions for $\bar s^{\rm TT}$ around $\Omega\simeq 15^\circ$ for $i <
90^\circ$ (and $\Omega \simeq 195^\circ$ for $i > 90^\circ$). The
performance of PSR~J1012+5307 is in accordance with its performance in
the $\hat\alpha_1$ test in~Ref.~\cite{sw12}, where PSR~J1738+0333 is
better in constraining the PF effects in the PPN formalism as
well. Partial reason for this performance will be discussed in the
next section.

The constraints from PSR~J0348+0432 are illustrated in
Figure~\ref{fig:J0348} as a function of $\Omega$. The joint limit
shows a smooth constraint versus $\Omega \in [0^\circ, 360^\circ)$,
  and from its joint constraint, we get the worst limit,
%---------------------------------------------------------------------
\begin{equation}
  |\bar s^{\rm TT}| < 7.2 \times 10^{-5}  \quad (95\%~{\rm CL}) \,,
\end{equation}
%---------------------------------------------------------------------
at $\Omega \simeq 270^\circ$ for $i < 90^\circ$ (and at $\Omega \simeq
90^\circ$ for $i > 90^\circ$). It is about five times weaker than the
limit from PSR~J1738+0333, yet still 100 times better than the GPB
limit.

%=====================================================================
\section{Discussions}\label{sec:discussion}
%=====================================================================

%---------------------------------------------------------------------
\begin{figure}
  \centering
  \includegraphics[width=9cm]{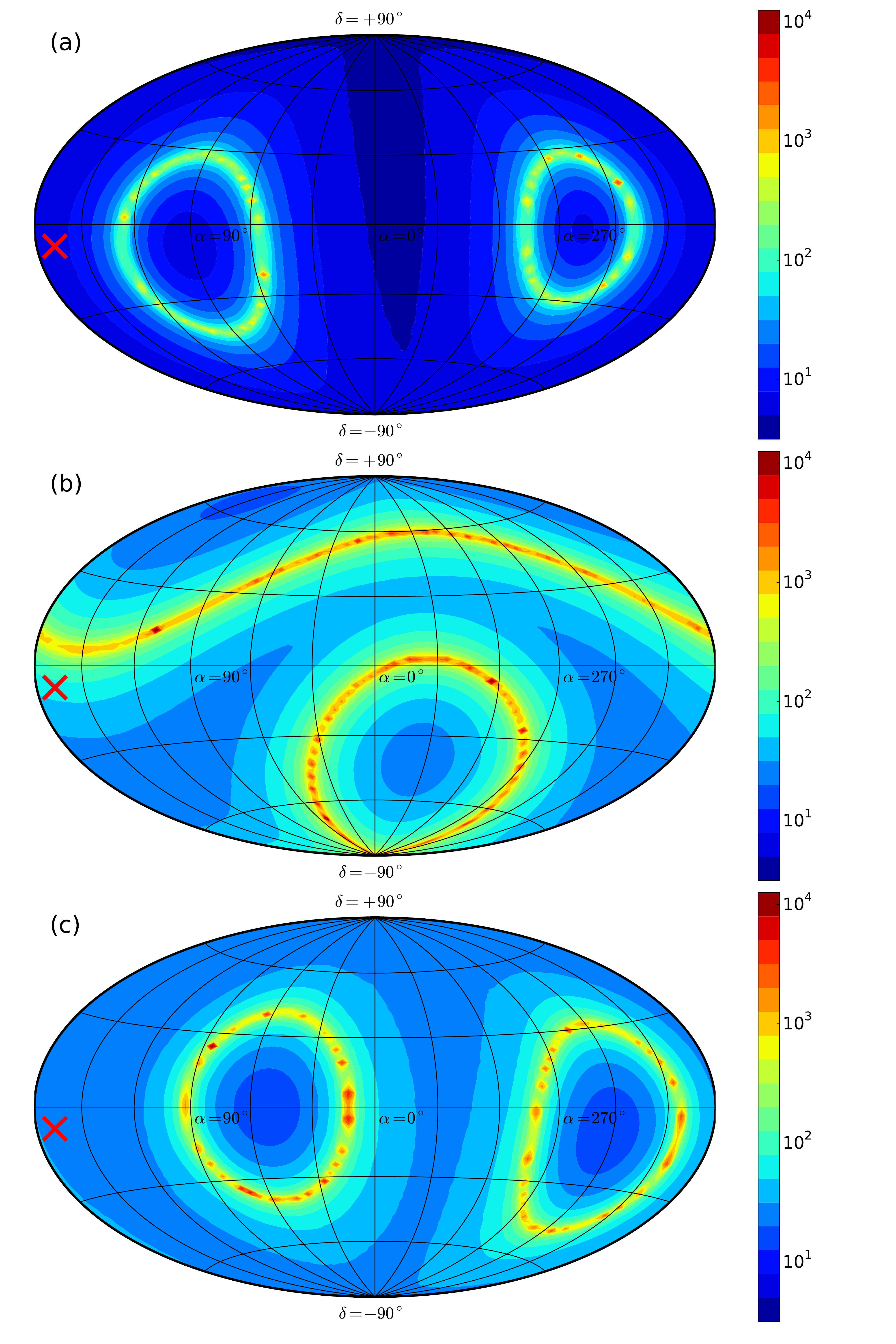}
  \caption{The constraints of $|\bar s^{\rm TT}|$ from
    PSRs~J1738+0333~(a), J1012+5307~(b), and J0348+0432~(c) for
    different PFs. Here $(\alpha,\delta)$ denotes the direction of the
    absolute motion of the Solar system with respect to a PF. The
    magnitude of this motion is assumed to be $|{\bf w}_\odot| \simeq
    369\,{\rm km\,s}^{-1}$. The red cross denotes the direction of the
    CMB frame, that has equatorial coordinates, $(\alpha,\delta)_{\rm
      CMB} \simeq (168^\circ,-7^\circ)$. Notice that, for convenience
    in comparison, all three figures use the same color scaling, as
    displayed on the right side. The unit for $\bar s^{\rm TT}$ is
    $10^{-6}$. \label{fig:sky3}}
\end{figure}
%---------------------------------------------------------------------

%---------------------------------------------------------------------
\begin{figure}
  \centering
  \includegraphics[width=9cm]{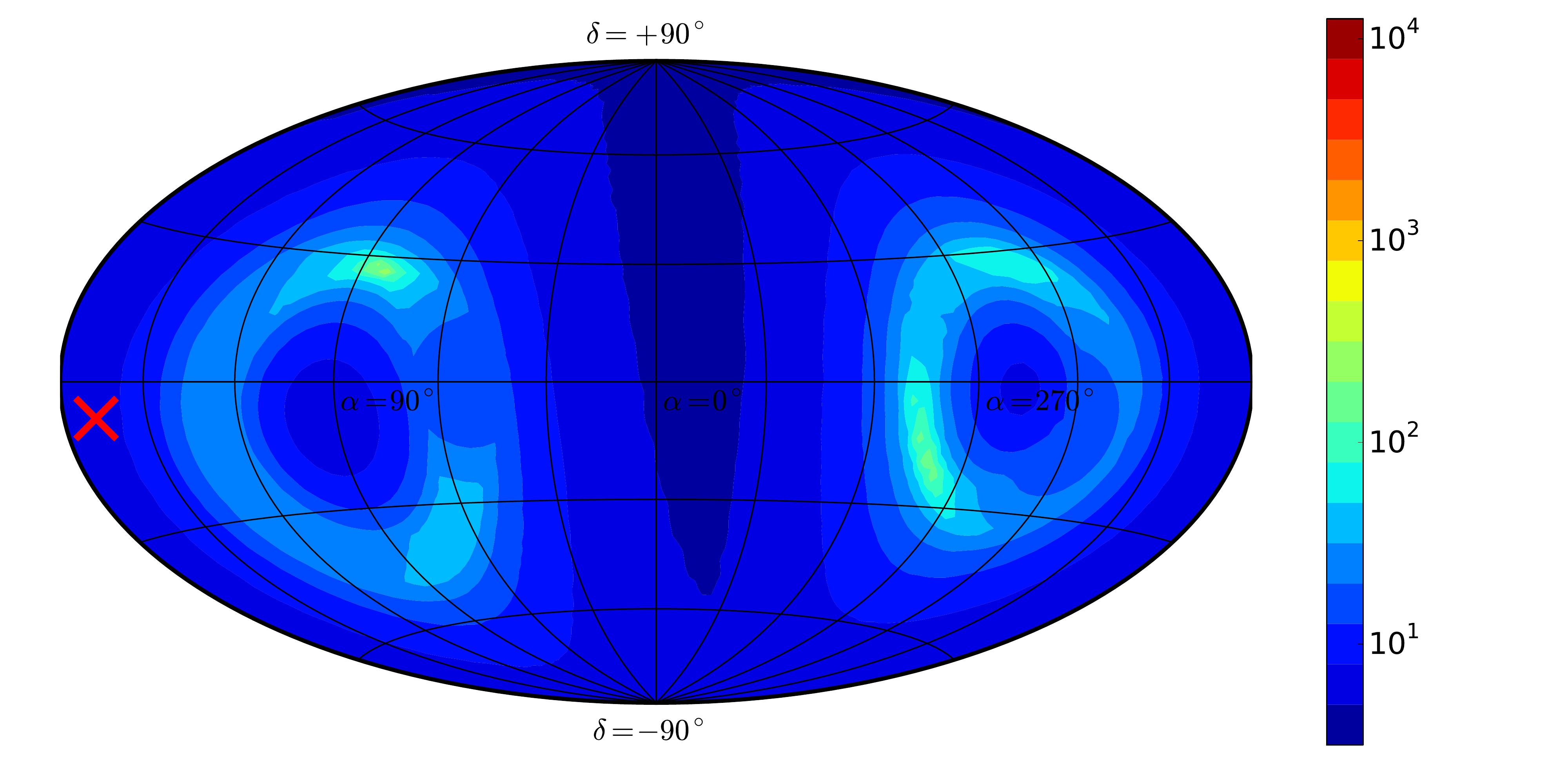}
  \caption{Same as Figure~\ref{fig:sky3}, for the combined constraint
    from three binary pulsars --- PSRs~J1738+0333, J1012+5307, and
    J0348+0432. For convenience in comparison, it uses the same color
    scaling as that of Figure~\ref{fig:sky3}. The unit for $\bar
    s^{\rm TT}$ is $10^{-6}$.\label{fig:sky1}}
\end{figure}
%---------------------------------------------------------------------

While using the isotropic CMB frame as the PF, we are basically
assuming that the PF is determined by the global matter distribution
in the Universe, and that the extra vectorial or tensorial components
of gravitational interaction are of long range, at least comparable to
the Hubble radius. When this is generally the most plausible
assumption, it is still interesting to consider other PFs, and further
show the power of binary pulsars in constraining LLI violation in the
gravity sector.

It is straightforward to apply the computations above to other PFs. We
here assume that the Solar system is moving with respect to a PF, with
a velocity, ${\bf w}_{\odot}$. Its magnitude is assumed to be $|{\bf
  w}_{\odot}| \simeq 369\,{\rm km\,s}^{-1}$, while its direction is
$(\alpha,\delta)$ when expressed in the equatorial coordinate.  For
every pair of $(\alpha,\delta)$, we perform the test above to get the
worst constraint of $\bar s^{\rm TT}$ from joint limits of $\dot x$,
$\dot\eta$, and $\dot\kappa$, for each pulsar. These limits at 68.3\%
CL are depicted as contours in Figure~\ref{fig:sky3} for
PSRs~J1738+0333, J1012+5307, and J0348+0432. For the reason of
computational burden, we have ignored the measurement uncertainties of
parameters except those of $\dot x$, $\dot\eta$, and $\dot\kappa$ for
the illustrated results in the figure. We have checked that this
treatment only affects our results at a level $\lesssim20\%$. We can
read out the following results from Figure~\ref{fig:sky3} ---
%---------------------------------------------------------------------
\begin{itemize}
\item For each pulsar, there are two cones with opposite directions of
  ${\bf w}_\odot$ that provide almost no constraint on $\bar s^{\rm
  TT}$ for the worst limit versus $\Omega$. This is similar to the PF
  tests in Ref.~\cite{sw12} (see their Figure 8).\footnote{Figure 8 is
    plotted in Galactic coordinates, while here we are using
    equatorial coordinates.}
\item With current timing precision, PSR~J1738+0333 has a better
  power, for most directions of ${\bf w}_\odot$, in constraining $\bar
  s^{\rm TT}$ over PSRs~J1012+5307 and J0348+0432.
\item The CMB frame has a direction $(\alpha,\delta)_{\rm CMB} \simeq
  (168^\circ,-7^\circ)$, denoted as a red cross in
  Figure~\ref{fig:sky3}. Compared with the other two pulsars, it is
  nearer to the ``unconstraining zones'' of PSR~J1012+5307, hence this
  pulsar provides a much worse constraint on $\bar s^{\rm TT}$ when
  the CMB frame is assumed to be the PF.
\end{itemize}
%---------------------------------------------------------------------

It is interesting to notice that, the ``unconstraining zones'' of
different pulsars are different. Therefore, when combining all
constraints from all pulsars can effectively eliminate these
``unconstraining zones''. The result of such a combination is shown in
Figure~\ref{fig:sky1}. It is clearly seen that such a combination
indeed eliminates almost all unconstrained directions. The worst
directions still provide tight constraints of $|\bar s^{\rm TT}|
\lesssim {\cal O}(10^{-4})$ at 68.3\% CL, where $|{\bf w}_\odot|$ is
again assumed to be 369\,km\,s$^{-1}$. For a different velocity of the
Solar system, the constraint of $\bar s^{\rm TT}$ can be obtained from
our result with a proper rescaling.\footnote{The rescaling is not a
  linear one, because it involves a vectorial superposition of the
  velocity of the pulsar system with respect to the Solar system and
  ${\bf w}_\odot$.}

Nevertheless, it is dangerous to naively combine limits from different
binary pulsars if strong-field effects, associated with NSs, play a
significant role.  In scalar-tensor theories, Damour and
Esposito-Far{\`e}se discovered that, within some parameter space, the
strong fields associated with NSs can develop nonperturbative
gravitational effects which can modify the orbital dynamics
significantly. This phenomenon is named as
``scalarization''~\cite{de93}. Although in the Einstein-\AE{}ther
theory and Ho{\v r}ava-Lifshitz gravity, no similar phenomena were
found~\cite{ybby14}, there is no formal proof yet that, for gravity
theories with LLI violation, similar strong-field dynamics like
``scalarization'' is absent in general with compact bodies. Therefore,
to be conservative, we claim that the limit of $\bar s^{\rm TT}$
obtained in this work should be treated as a strong-field limit to its
weak-field counterpart. This statement does not necessarily mean that
the limit from binary pulsars is ``weak'', on the contrary, in most
cases the limits from strongly self-gravitating bodies are {\it much}
stronger than the corresponding limits from weakly gravitating
bodies. For example, the parameter for the strong equivalence
principle violation, $\Delta$, obtained from binary pulsars (see
{\it e.g.} Refs.~\cite{sta03,sfl+05,wex14}), is in general a factor of
${\cal C}_{\rm NS} / {\cal C}_0$ stronger in terms of the Nordtvedt
parameter, than a corresponding limit from a weakly gravitating body
with compactness ${\cal C}_0$. For NSs, we have ${\cal C}_{\rm NS}
\sim 0.2$, while for the Earth and the Sun, we have ${\cal C}_\oplus
\sim 10^{-10}$ and ${\cal C}_\odot \sim 10^{-6}$ respectively.

In scalar-tensor theories, if the mentioned scalarization happens, the
strong-field parameters of gravity theories, such as the PPN
parameters, $\beta$ and $\gamma$, will depend on the compactness of
the gravitating body. In other words, they become system dependent. If
such phenomenon also happens here, we will expect different values of
$\bar s^{\rm TT}$ for PSRs~J1738+0333, J1012+5307, and
J0348+0432. Nonetheless, if the dynamics is still within the
perturbative region of the gravity theory, these $\bar s^{\rm TT}$'s
will be of similar values, and in this situation, we are eligible to
combine results from different binary pulsars, as done in
Figure~\ref{fig:sky1}.

Worthy to reemphasize that the $\bar s^{\rm TT}$ constraints here are
based on robust designs of tests, in contrast to the probabilistic
tests performed in Ref.~\cite{shao14}. The achievement is made
possible with multiple observables measured simultaneously. Even
without a clear knowledge on $\Omega$, picking the worst limit out of
all possible limits makes the test very robust. Such ideas were also
proposed in Refs.~\cite{fkw12,sw12} under different topics on tests of
gravity. In the future if we can measure the longitude of ascending
node for these binary pulsars, with {\it e.g.} interstellar
scintillation~\cite{rcn+14}, we will gain further in constraining the
SME parameters.

In the pure-gravity sector of mSME, there are in total nine degrees of
freedom to deviate away from GR at leading order~\cite{bk06}. Here to
focus on the theme of testing $\bar s^{\rm TT}$ and to reduce the
burden of work, we have assumed that there exists a PF where the
spacetime is isotropic. Universal anisotropy can exist in principle
within the framework of SME.  Therefore, the actual constraint in this
paper could be a linear combination of $\bar s^{\rm TT}$ and the other
time-spatial and spatial-spatial components of $\bar s^{\mu\nu}$.
However, it was already shown in Ref.~\cite{shao14}, that the
time-spatial and spatial-spatial components of $\bar s^{\mu\nu}$ were
constrained to very small values at levels of ${\cal
  O}(10^{-9})$--${\cal O}(10^{-11})$, and more importantly, with
multiple pulsar systems with different positions in the sky, different
spatial orientations of orbits, and different 3-dimensional systematic
velocities, the limits of these components tend to have very little
mutual correlations (see Figure 3 in Ref.~\cite{shao14}). Therefore,
the limit obtained in this paper is reliable even when there exists no
PF at all for the gravity sector in SME.

From another viewpoint, in the past few years, the time-spatial and
spatial-spatial components in the Sun-centered celestial-equatorial
frame were already stringently constrained by lunar laser
ranging~\cite{bcs07}, atom interferometry~\cite{mch+08,cch+09}, and
radio pulsars~\cite{shao14}.  Among these, the current best limits
were obtained from the systematic analysis of 27 tests from 13 pulsar
systems~\cite{shao14}. From the joint analysis of orbital dynamics of
binary pulsars and spin evolution of solitary pulsars, the components
of $\bar s^{\mu\nu}$ were constrained to be, at 68\% confidence level,
$\lesssim 10^{-9}$ for $\bar s^{\rm TX}$, $\bar s^{\rm TY}$, and $\bar
s^{\rm TZ}$, and $\lesssim 10^{-11}$ for $\bar s^{\rm XY}$, $\bar
s^{\rm YZ}$, $\bar s^{\rm XZ}$, $\bar s^{\rm XX} - \bar s^{\rm YY}$,
and $\bar s^{\rm XX} + \bar s^{\rm YY} - 2 \bar s^{\rm ZZ}$ (see Table
1 in Ref.~\cite{shao14}).  Therefore, we can {\it empirically} write
the $\bar s^{\mu\nu}$ field in the Sun-centered frame as
%---------------------------------------------------------------------
\begin{equation}
  \bar s^{\mu\nu} \simeq  \left(
  \begin{array}{cccc}
    \bar s^{\rm TT} & 0 & 0 & 0 \\
    0 & \frac{1}{3}\bar s^{\rm TT} & 0 & 0 \\
    0 & 0 & \frac{1}{3}\bar s^{\rm TT} & 0 \\
    0 & 0 & 0 & \frac{1}{3}\bar s^{\rm TT} \\
  \end{array}
  \right) \,.
\end{equation}
%---------------------------------------------------------------------
By writing down the above {\it numerical} expression with possibly
dominant nonzero components from purely empirical
evidence~\cite{bcs07,mch+08,cch+09,shao14}, we are free of assuming
the existence of a PF. The calculation to constrain such an $\bar
s^{\mu\nu}$ is straightforward with the analysis in this paper. With
Monte Carlo simulations properly accounting for all measurement
uncertainties, we found that the best robust limit of $\bar s^{\rm
  TT}$ still comes from PSR~J1738+0333, that gives
%---------------------------------------------------------------------
\begin{equation}
  |\bar s^{\rm TT}| < 2.8 \times 10^{-4}  \quad (95\%~{\rm CL})\,.
\end{equation}
%---------------------------------------------------------------------
It is weaker than the limit with the isotropic frame of CMB as the PF,
due to the fact that the boost between PSR~J1738+0333 and the Solar
system is only about $|{\bf v}| = 74.8 \pm 9.5\,{\rm km\,s}^{-1}$. The
systematic velocity is related to the evolutionary history of NS-WD
systems~\cite{lk04}. Nevertheless, this limit, free of the assumption
that there exists a PF, is still one order of magnitude better than
the current best limit.

%=====================================================================
\section{Summary}\label{sec:summary}
%=====================================================================

In this paper, we propose a new idea to test the $\bar s^{\rm TT}$
component in the pure-gravity sector of mSME by utilizing the boost
between different frames.  A new robust limit, in the standard
Sun-centered equatorial-celestial coordinate frame, is obtained from
PSR~J1738+0333,
%---------------------------------------------------------------------
\begin{equation}
  |\bar s^{\rm TT}| < 1.6 \times 10^{-5}  \quad (95\%~{\rm CL})\,,
\end{equation}
%---------------------------------------------------------------------
when the isotropic CMB frame is assumed to be the PF. The limit is
about 500 times better than the current best limit from Gravity Probe
B~\cite{beo13}.

The idea of mixing different components in the condensed (cosmic or
even local) tensor fields with a full Lorentz transformation is also
applicable in other sectors of SME with careful studies, as
demonstrated in Refs.~\cite{cbp+04,hac+08,gkv14}. Although such a
boost is usually quite small ({\it e.g.} ${\cal O}(10^{-3})$ for
binary pulsars), with some astrophysical systems, the method could
become useful with precision experiments, as done here with the
state-of-the-art pulsar timing experiments.

%=====================================================================
\section*{Acknowledgements}
%=====================================================================

We thank Alan Kosteleck{\'y} for carefully reading the manuscript and
kindly providing useful comments. We are grateful to Quentin Bailey,
Paulo Freire, Jay Tasson, and Norbert Wex for helpful discussions.

%=====================================================================

%\bibliography{./../reference}{}

\begin{thebibliography}{54}
\expandafter\ifx\csname natexlab\endcsname\relax\def\natexlab#1{#1}\fi
\expandafter\ifx\csname bibnamefont\endcsname\relax
  \def\bibnamefont#1{#1}\fi
\expandafter\ifx\csname bibfnamefont\endcsname\relax
  \def\bibfnamefont#1{#1}\fi
\expandafter\ifx\csname citenamefont\endcsname\relax
  \def\citenamefont#1{#1}\fi
\expandafter\ifx\csname url\endcsname\relax
  \def\url#1{\texttt{#1}}\fi
\expandafter\ifx\csname urlprefix\endcsname\relax\def\urlprefix{URL }\fi
\providecommand{\bibinfo}[2]{#2}
\providecommand{\eprint}[2][]{\url{#2}}

\bibitem[{\citenamefont{{Misner} et~al.}(1973)\citenamefont{{Misner}, {Thorne},
  and {Wheeler}}}]{mtw73}
\bibinfo{author}{\bibfnamefont{C.~W.} \bibnamefont{{Misner}}},
  \bibinfo{author}{\bibfnamefont{K.~S.} \bibnamefont{{Thorne}}},
  \bibnamefont{and} \bibinfo{author}{\bibfnamefont{J.~A.}
  \bibnamefont{{Wheeler}}}, \emph{\bibinfo{title}{{Gravitation}}}
  (\bibinfo{publisher}{San Francisco: W.H.~Freeman and Company},
  \bibinfo{year}{1973}).

\bibitem[{\citenamefont{{Will}}(2014)}]{will14}
\bibinfo{author}{\bibfnamefont{C.~M.} \bibnamefont{{Will}}},
  \bibinfo{journal}{Living Reviews in Relativity}
  \textbf{\bibinfo{volume}{17}}, \bibinfo{pages}{4} (\bibinfo{year}{2014}),
  \eprint{arXiv:1403.7377}.

\bibitem[{\citenamefont{{Einstein}}(1916)}]{ein16}
\bibinfo{author}{\bibfnamefont{A.}~\bibnamefont{{Einstein}}},
  \bibinfo{journal}{Annalen der Physik} \textbf{\bibinfo{volume}{354}},
  \bibinfo{pages}{769} (\bibinfo{year}{1916}).

\bibitem[{\citenamefont{{Shapiro}}(1964)}]{sha64}
\bibinfo{author}{\bibfnamefont{I.~I.} \bibnamefont{{Shapiro}}},
  \bibinfo{journal}{Physical Review Letters} \textbf{\bibinfo{volume}{13}},
  \bibinfo{pages}{789} (\bibinfo{year}{1964}).

\bibitem[{\citenamefont{{Matthews} and {Sandage}}(1963)}]{ms63}
\bibinfo{author}{\bibfnamefont{T.~A.} \bibnamefont{{Matthews}}}
  \bibnamefont{and} \bibinfo{author}{\bibfnamefont{A.~R.}
  \bibnamefont{{Sandage}}}, \bibinfo{journal}{\apj}
  \textbf{\bibinfo{volume}{138}}, \bibinfo{pages}{30} (\bibinfo{year}{1963}).

\bibitem[{\citenamefont{{Hewish} et~al.}(1968)\citenamefont{{Hewish}, {Bell},
  {Pilkington}, {Scott}, and {Collins}}}]{hbp+68}
\bibinfo{author}{\bibfnamefont{A.}~\bibnamefont{{Hewish}}},
  \bibinfo{author}{\bibfnamefont{S.~J.} \bibnamefont{{Bell}}},
  \bibinfo{author}{\bibfnamefont{J.~D.~H.} \bibnamefont{{Pilkington}}},
  \bibinfo{author}{\bibfnamefont{P.~F.} \bibnamefont{{Scott}}},
  \bibnamefont{and} \bibinfo{author}{\bibfnamefont{R.~A.}
  \bibnamefont{{Collins}}}, \bibinfo{journal}{\nat}
  \textbf{\bibinfo{volume}{217}}, \bibinfo{pages}{709} (\bibinfo{year}{1968}).

\bibitem[{\citenamefont{{Penzias} and {Wilson}}(1965)}]{pw65}
\bibinfo{author}{\bibfnamefont{A.~A.} \bibnamefont{{Penzias}}}
  \bibnamefont{and} \bibinfo{author}{\bibfnamefont{R.~W.}
  \bibnamefont{{Wilson}}}, \bibinfo{journal}{\apj}
  \textbf{\bibinfo{volume}{142}}, \bibinfo{pages}{419} (\bibinfo{year}{1965}).

\bibitem[{\citenamefont{{Everitt} et~al.}(2011)\citenamefont{{Everitt},
  {Debra}, {Parkinson}, {Turneaure}, {Conklin}, {Heifetz}, {Keiser},
  {Silbergleit}, {Holmes}, {Kolodziejczak} et~al.}}]{edp+11}
\bibinfo{author}{\bibfnamefont{C.~W.~F.} \bibnamefont{{Everitt}}},
  \bibinfo{author}{\bibfnamefont{D.~B.} \bibnamefont{{Debra}}},
  \bibinfo{author}{\bibfnamefont{B.~W.} \bibnamefont{{Parkinson}}},
  \bibinfo{author}{\bibfnamefont{J.~P.} \bibnamefont{{Turneaure}}},
  \bibinfo{author}{\bibfnamefont{J.~W.} \bibnamefont{{Conklin}}},
  \bibinfo{author}{\bibfnamefont{M.~I.} \bibnamefont{{Heifetz}}},
  \bibinfo{author}{\bibfnamefont{G.~M.} \bibnamefont{{Keiser}}},
  \bibinfo{author}{\bibfnamefont{A.~S.} \bibnamefont{{Silbergleit}}},
  \bibinfo{author}{\bibfnamefont{T.}~\bibnamefont{{Holmes}}},
  \bibinfo{author}{\bibfnamefont{J.}~\bibnamefont{{Kolodziejczak}}},
  \bibnamefont{et~al.}, \bibinfo{journal}{Physical Review Letters}
  \textbf{\bibinfo{volume}{106}}, \bibinfo{eid}{221101} (\bibinfo{year}{2011}),
  \eprint{arXiv:1105.3456}.

\bibitem[{\citenamefont{{Kramer} et~al.}(2006)\citenamefont{{Kramer}, {Stairs},
  {Manchester}, {McLaughlin}, {Lyne}, {Ferdman}, {Burgay}, {Lorimer},
  {Possenti}, {D'Amico} et~al.}}]{ksm+06}
\bibinfo{author}{\bibfnamefont{M.}~\bibnamefont{{Kramer}}},
  \bibinfo{author}{\bibfnamefont{I.~H.} \bibnamefont{{Stairs}}},
  \bibinfo{author}{\bibfnamefont{R.~N.} \bibnamefont{{Manchester}}},
  \bibinfo{author}{\bibfnamefont{M.~A.} \bibnamefont{{McLaughlin}}},
  \bibinfo{author}{\bibfnamefont{A.~G.} \bibnamefont{{Lyne}}},
  \bibinfo{author}{\bibfnamefont{R.~D.} \bibnamefont{{Ferdman}}},
  \bibinfo{author}{\bibfnamefont{M.}~\bibnamefont{{Burgay}}},
  \bibinfo{author}{\bibfnamefont{D.~R.} \bibnamefont{{Lorimer}}},
  \bibinfo{author}{\bibfnamefont{A.}~\bibnamefont{{Possenti}}},
  \bibinfo{author}{\bibfnamefont{N.}~\bibnamefont{{D'Amico}}},
  \bibnamefont{et~al.}, \bibinfo{journal}{Science}
  \textbf{\bibinfo{volume}{314}}, \bibinfo{pages}{97} (\bibinfo{year}{2006}),
  \eprint{arXiv:astro-ph/0609417}.

\bibitem[{\citenamefont{{Breton} et~al.}(2008)\citenamefont{{Breton}, {Kaspi},
  {Kramer}, {McLaughlin}, {Lyutikov}, {Ransom}, {Stairs}, {Ferdman}, {Camilo},
  and {Possenti}}}]{bkk+08}
\bibinfo{author}{\bibfnamefont{R.~P.} \bibnamefont{{Breton}}},
  \bibinfo{author}{\bibfnamefont{V.~M.} \bibnamefont{{Kaspi}}},
  \bibinfo{author}{\bibfnamefont{M.}~\bibnamefont{{Kramer}}},
  \bibinfo{author}{\bibfnamefont{M.~A.} \bibnamefont{{McLaughlin}}},
  \bibinfo{author}{\bibfnamefont{M.}~\bibnamefont{{Lyutikov}}},
  \bibinfo{author}{\bibfnamefont{S.~M.} \bibnamefont{{Ransom}}},
  \bibinfo{author}{\bibfnamefont{I.~H.} \bibnamefont{{Stairs}}},
  \bibinfo{author}{\bibfnamefont{R.~D.} \bibnamefont{{Ferdman}}},
  \bibinfo{author}{\bibfnamefont{F.}~\bibnamefont{{Camilo}}}, \bibnamefont{and}
  \bibinfo{author}{\bibfnamefont{A.}~\bibnamefont{{Possenti}}},
  \bibinfo{journal}{Science} \textbf{\bibinfo{volume}{321}},
  \bibinfo{pages}{104} (\bibinfo{year}{2008}), \eprint{arXiv:0807.2644}.

\bibitem[{\citenamefont{{Sathyaprakash} and {Schutz}}(2009)}]{ss09}
\bibinfo{author}{\bibfnamefont{B.~S.} \bibnamefont{{Sathyaprakash}}}
  \bibnamefont{and} \bibinfo{author}{\bibfnamefont{B.~F.}
  \bibnamefont{{Schutz}}}, \bibinfo{journal}{Living Reviews in Relativity}
  \textbf{\bibinfo{volume}{12}}, \bibinfo{pages}{2} (\bibinfo{year}{2009}),
  \eprint{arXiv:0903.0338}.

\bibitem[{\citenamefont{{Pitkin} et~al.}(2011)\citenamefont{{Pitkin}, {Reid},
  {Rowan}, and {Hough}}}]{prrh11}
\bibinfo{author}{\bibfnamefont{M.}~\bibnamefont{{Pitkin}}},
  \bibinfo{author}{\bibfnamefont{S.}~\bibnamefont{{Reid}}},
  \bibinfo{author}{\bibfnamefont{S.}~\bibnamefont{{Rowan}}}, \bibnamefont{and}
  \bibinfo{author}{\bibfnamefont{J.}~\bibnamefont{{Hough}}},
  \bibinfo{journal}{Living Reviews in Relativity}
  \textbf{\bibinfo{volume}{14}}, \bibinfo{pages}{5} (\bibinfo{year}{2011}),
  \eprint{arXiv:1102.3355}.

\bibitem[{\citenamefont{{Gair} et~al.}(2013)\citenamefont{{Gair}, {Vallisneri},
  {Larson}, and {Baker}}}]{gvlb13}
\bibinfo{author}{\bibfnamefont{J.~R.} \bibnamefont{{Gair}}},
  \bibinfo{author}{\bibfnamefont{M.}~\bibnamefont{{Vallisneri}}},
  \bibinfo{author}{\bibfnamefont{S.~L.} \bibnamefont{{Larson}}},
  \bibnamefont{and} \bibinfo{author}{\bibfnamefont{J.~G.}
  \bibnamefont{{Baker}}}, \bibinfo{journal}{Living Reviews in Relativity}
  \textbf{\bibinfo{volume}{16}}, \bibinfo{pages}{7} (\bibinfo{year}{2013}),
  \eprint{arXiv:1212.5575}.

\bibitem[{\citenamefont{{Hobbs}}(2013)}]{hob13}
\bibinfo{author}{\bibfnamefont{G.}~\bibnamefont{{Hobbs}}},
  \bibinfo{journal}{Classical and Quantum Gravity}
  \textbf{\bibinfo{volume}{30}}, \bibinfo{eid}{224007} (\bibinfo{year}{2013}),
  \eprint{arXiv:1307.2629}.

\bibitem[{\citenamefont{{Kramer} and {Champion}}(2013)}]{kc13}
\bibinfo{author}{\bibfnamefont{M.}~\bibnamefont{{Kramer}}} \bibnamefont{and}
  \bibinfo{author}{\bibfnamefont{D.~J.} \bibnamefont{{Champion}}},
  \bibinfo{journal}{Classical and Quantum Gravity}
  \textbf{\bibinfo{volume}{30}}, \bibinfo{eid}{224009} (\bibinfo{year}{2013}).

\bibitem[{\citenamefont{{McLaughlin}}(2013)}]{mcl13}
\bibinfo{author}{\bibfnamefont{M.~A.} \bibnamefont{{McLaughlin}}},
  \bibinfo{journal}{Classical and Quantum Gravity}
  \textbf{\bibinfo{volume}{30}}, \bibinfo{eid}{224008} (\bibinfo{year}{2013}),
  \eprint{arXiv:1310.0758}.

\bibitem[{\citenamefont{{Yunes} and {Siemens}}(2013)}]{ys13}
\bibinfo{author}{\bibfnamefont{N.}~\bibnamefont{{Yunes}}} \bibnamefont{and}
  \bibinfo{author}{\bibfnamefont{X.}~\bibnamefont{{Siemens}}},
  \bibinfo{journal}{Living Reviews in Relativity}
  \textbf{\bibinfo{volume}{16}}, \bibinfo{pages}{9} (\bibinfo{year}{2013}),
  \eprint{arXiv:1304.3473}.

\bibitem[{\citenamefont{{Capozziello} and {de Laurentis}}(2011)}]{cl11}
\bibinfo{author}{\bibfnamefont{S.}~\bibnamefont{{Capozziello}}}
  \bibnamefont{and} \bibinfo{author}{\bibfnamefont{M.}~\bibnamefont{{de
  Laurentis}}}, \bibinfo{journal}{Phys. Rep.} \textbf{\bibinfo{volume}{509}},
  \bibinfo{pages}{167} (\bibinfo{year}{2011}), \eprint{arXiv:1108.6266}.

\bibitem[{\citenamefont{{Clifton} et~al.}(2012)\citenamefont{{Clifton},
  {Ferreira}, {Padilla}, and {Skordis}}}]{cfps12}
\bibinfo{author}{\bibfnamefont{T.}~\bibnamefont{{Clifton}}},
  \bibinfo{author}{\bibfnamefont{P.~G.} \bibnamefont{{Ferreira}}},
  \bibinfo{author}{\bibfnamefont{A.}~\bibnamefont{{Padilla}}},
  \bibnamefont{and}
  \bibinfo{author}{\bibfnamefont{C.}~\bibnamefont{{Skordis}}},
  \bibinfo{journal}{Phys. Rep.} \textbf{\bibinfo{volume}{513}},
  \bibinfo{pages}{1} (\bibinfo{year}{2012}), \eprint{arXiv:1106.2476}.

\bibitem[{\citenamefont{{Ashby}}(2003)}]{ash03}
\bibinfo{author}{\bibfnamefont{N.}~\bibnamefont{{Ashby}}},
  \bibinfo{journal}{Living Reviews in Relativity} \textbf{\bibinfo{volume}{6}},
  \bibinfo{pages}{1} (\bibinfo{year}{2003}).

\bibitem[{\citenamefont{{Abramowicz} and {Fragile}}(2013)}]{af13}
\bibinfo{author}{\bibfnamefont{M.~A.} \bibnamefont{{Abramowicz}}}
  \bibnamefont{and} \bibinfo{author}{\bibfnamefont{P.~C.}
  \bibnamefont{{Fragile}}}, \bibinfo{journal}{Living Reviews in Relativity}
  \textbf{\bibinfo{volume}{16}}, \bibinfo{pages}{1} (\bibinfo{year}{2013}),
  \eprint{arXiv:1104.5499}.

\bibitem[{\citenamefont{{Ho{\v r}ava}}(2009)}]{hor09}
\bibinfo{author}{\bibfnamefont{P.}~\bibnamefont{{Ho{\v r}ava}}},
  \bibinfo{journal}{\prd} \textbf{\bibinfo{volume}{79}}, \bibinfo{eid}{084008}
  (\bibinfo{year}{2009}), \eprint{arXiv:0901.3775}.

\bibitem[{\citenamefont{{Jacobson} and {Mattingly}}(2001)}]{jm01}
\bibinfo{author}{\bibfnamefont{T.}~\bibnamefont{{Jacobson}}} \bibnamefont{and}
  \bibinfo{author}{\bibfnamefont{D.}~\bibnamefont{{Mattingly}}},
  \bibinfo{journal}{\prd} \textbf{\bibinfo{volume}{64}}, \bibinfo{eid}{024028}
  (\bibinfo{year}{2001}), \eprint{arXiv:gr-qc/0007031}.

\bibitem[{\citenamefont{{Bailey} and {Kosteleck{\'y}}}(2006)}]{bk06}
\bibinfo{author}{\bibfnamefont{Q.~G.} \bibnamefont{{Bailey}}} \bibnamefont{and}
  \bibinfo{author}{\bibfnamefont{V.~A.} \bibnamefont{{Kosteleck{\'y}}}},
  \bibinfo{journal}{\prd} \textbf{\bibinfo{volume}{74}}, \bibinfo{eid}{045001}
  (\bibinfo{year}{2006}), \eprint{arXiv:gr-qc/0603030}.

\bibitem[{\citenamefont{{Bailey} et~al.}(2013)\citenamefont{{Bailey},
  {Everett}, and {Overduin}}}]{beo13}
\bibinfo{author}{\bibfnamefont{Q.~G.} \bibnamefont{{Bailey}}},
  \bibinfo{author}{\bibfnamefont{R.~D.} \bibnamefont{{Everett}}},
  \bibnamefont{and} \bibinfo{author}{\bibfnamefont{J.~M.}
  \bibnamefont{{Overduin}}}, \bibinfo{journal}{\prd}
  \textbf{\bibinfo{volume}{88}}, \bibinfo{eid}{102001} (\bibinfo{year}{2013}),
  \eprint{arXiv:1309.6399}.

\bibitem[{\citenamefont{{Shao}}(2014)}]{shao14}
\bibinfo{author}{\bibfnamefont{L.}~\bibnamefont{{Shao}}},
  \bibinfo{journal}{Physical Review Letters} \textbf{\bibinfo{volume}{112}},
  \bibinfo{eid}{111103} (\bibinfo{year}{2014}), \eprint{arXiv:1402.6452}.

\bibitem[{\citenamefont{{Kosteleck{\'y}} and
  {Samuel}}(1989{\natexlab{a}})}]{ks89a}
\bibinfo{author}{\bibfnamefont{V.~A.} \bibnamefont{{Kosteleck{\'y}}}}
  \bibnamefont{and} \bibinfo{author}{\bibfnamefont{S.}~\bibnamefont{{Samuel}}},
  \bibinfo{journal}{\prd} \textbf{\bibinfo{volume}{39}}, \bibinfo{pages}{683}
  (\bibinfo{year}{1989}{\natexlab{a}}).

\bibitem[{\citenamefont{{Kosteleck{\'y}} and
  {Samuel}}(1989{\natexlab{b}})}]{ks89b}
\bibinfo{author}{\bibfnamefont{V.~A.} \bibnamefont{{Kosteleck{\'y}}}}
  \bibnamefont{and} \bibinfo{author}{\bibfnamefont{S.}~\bibnamefont{{Samuel}}},
  \bibinfo{journal}{\prd} \textbf{\bibinfo{volume}{40}}, \bibinfo{pages}{1886}
  (\bibinfo{year}{1989}{\natexlab{b}}).

\bibitem[{\citenamefont{{Colladay} and {Kosteleck{\'y}}}(1997)}]{ck97}
\bibinfo{author}{\bibfnamefont{D.}~\bibnamefont{{Colladay}}} \bibnamefont{and}
  \bibinfo{author}{\bibfnamefont{V.~A.} \bibnamefont{{Kosteleck{\'y}}}},
  \bibinfo{journal}{\prd} \textbf{\bibinfo{volume}{55}}, \bibinfo{pages}{6760}
  (\bibinfo{year}{1997}), \eprint{arXiv:hep-ph/9703464}.

\bibitem[{\citenamefont{{Colladay} and {Kosteleck{\'y}}}(1998)}]{ck98}
\bibinfo{author}{\bibfnamefont{D.}~\bibnamefont{{Colladay}}} \bibnamefont{and}
  \bibinfo{author}{\bibfnamefont{V.~A.} \bibnamefont{{Kosteleck{\'y}}}},
  \bibinfo{journal}{\prd} \textbf{\bibinfo{volume}{58}}, \bibinfo{eid}{116002}
  (\bibinfo{year}{1998}), \eprint{arXiv:hep-ph/9809521}.

\bibitem{nac69}
  O.~Nachtmann, %``CP violation and cosmological fields,''
  Conf.\ Proc.\ C {\bf 690224}, 485 (1969).
  %%CITATION = CONFP,C690224,485;%%

\bibitem{cn83}
  S.~Chadha and H.~B.~Nielsen,
  %``Lorentz invariance as a low-energy phenomenon,''
  Nucl.\ Phys.\ B {\bf 217}, 125 (1983).
  %%CITATION =NUPHA,B217,125;%%

\bibitem{kli00}
  F.~R.~Klinkhamer,
  %``A CPT anomaly,''
  Nucl.\ Phys.\ B {\bf 578}, 277 (2000) [arXiv:hep-th/9912169].
  %%CITATION = HEP-TH/9912169;%%
  
\bibitem[{\citenamefont{{Kosteleck{\'y}} and {Russell}}(2011)}]{kr11}
\bibinfo{author}{\bibfnamefont{V.~A.} \bibnamefont{{Kosteleck{\'y}}}}
  \bibnamefont{and}
  \bibinfo{author}{\bibfnamefont{N.}~\bibnamefont{{Russell}}},
  \bibinfo{journal}{Reviews of Modern Physics} \textbf{\bibinfo{volume}{83}},
  \bibinfo{pages}{11} (\bibinfo{year}{2011}), \eprint{arXiv:0801.0287}.

\bibitem[{\citenamefont{{Kosteleck{\'y}}}(2004)}]{kos04}
\bibinfo{author}{\bibfnamefont{V.~A.} \bibnamefont{{Kosteleck{\'y}}}},
  \bibinfo{journal}{\prd} \textbf{\bibinfo{volume}{69}}, \bibinfo{eid}{105009}
  (\bibinfo{year}{2004}), \eprint{arXiv:hep-th/0312310}.

\bibitem{bkx14} 
  Q.~G.~Bailey, V.~A.~Kosteleck{\' y}, and R.~Xu,
  arXiv:1410.6162.

\bibitem[{\citenamefont{{Kosteleck{\'y}} and {Tasson}}(2011)}]{kt11}
\bibinfo{author}{\bibfnamefont{V.~A.} \bibnamefont{{Kosteleck{\'y}}}}
  \bibnamefont{and} \bibinfo{author}{\bibfnamefont{J.~D.}
  \bibnamefont{{Tasson}}}, \bibinfo{journal}{\prd}
  \textbf{\bibinfo{volume}{83}}, \bibinfo{eid}{016013} (\bibinfo{year}{2011}),
  \eprint{arXiv:1006.4106}.

\bibitem[{\citenamefont{Higgs}(2014)}]{hig14}
\bibinfo{author}{\bibfnamefont{P.~W.} \bibnamefont{Higgs}},
  \bibinfo{journal}{Rev. Mod. Phys.} \textbf{\bibinfo{volume}{86}},
  \bibinfo{pages}{851} (\bibinfo{year}{2014}).

\bibitem[{\citenamefont{Englert}(2014)}]{eng14}
\bibinfo{author}{\bibfnamefont{F.}~\bibnamefont{Englert}},
  \bibinfo{journal}{Rev. Mod. Phys.} \textbf{\bibinfo{volume}{86}},
  \bibinfo{pages}{843} (\bibinfo{year}{2014}).


\bibitem[{\citenamefont{{Lorimer} and {Kramer}}(2004)}]{lk04}
\bibinfo{author}{\bibfnamefont{D.~R.} \bibnamefont{{Lorimer}}}
  \bibnamefont{and} \bibinfo{author}{\bibfnamefont{M.}~\bibnamefont{{Kramer}}},
  \emph{\bibinfo{title}{{Handbook of Pulsar Astronomy}}}
  (\bibinfo{publisher}{{Cambridge University Press}}, \bibinfo{year}{2004}).


\bibitem[{\citenamefont{{Lange} et~al.}(2001)\citenamefont{{Lange}, {Camilo},
  {Wex}, {Kramer}, {Backer}, {Lyne}, and {Doroshenko}}}]{lcw+01}
\bibinfo{author}{\bibfnamefont{C.}~\bibnamefont{{Lange}}},
  \bibinfo{author}{\bibfnamefont{F.}~\bibnamefont{{Camilo}}},
  \bibinfo{author}{\bibfnamefont{N.}~\bibnamefont{{Wex}}},
  \bibinfo{author}{\bibfnamefont{M.}~\bibnamefont{{Kramer}}},
  \bibinfo{author}{\bibfnamefont{D.~C.} \bibnamefont{{Backer}}},
  \bibinfo{author}{\bibfnamefont{A.~G.} \bibnamefont{{Lyne}}},
  \bibnamefont{and}
  \bibinfo{author}{\bibfnamefont{O.}~\bibnamefont{{Doroshenko}}},
  \bibinfo{journal}{\mnras} \textbf{\bibinfo{volume}{326}},
  \bibinfo{pages}{274} (\bibinfo{year}{2001}), \eprint{arXiv:astro-ph/0102309}.

\bibitem[{\citenamefont{{Will} and {Nordtvedt}}(1972)}]{wn72}
\bibinfo{author}{\bibfnamefont{C.~M.} \bibnamefont{{Will}}} \bibnamefont{and}
  \bibinfo{author}{\bibfnamefont{K.}~\bibnamefont{{Nordtvedt}},
  \bibfnamefont{Jr.}}, \bibinfo{journal}{\apj} \textbf{\bibinfo{volume}{177}},
  \bibinfo{pages}{757} (\bibinfo{year}{1972}).

\bibitem[{\citenamefont{{Nordtvedt} and {Will}}(1972)}]{nw72}
\bibinfo{author}{\bibfnamefont{K.}~\bibnamefont{{Nordtvedt}},
  \bibfnamefont{Jr.}} \bibnamefont{and} \bibinfo{author}{\bibfnamefont{C.~M.}
  \bibnamefont{{Will}}}, \bibinfo{journal}{\apj}
  \textbf{\bibinfo{volume}{177}}, \bibinfo{pages}{775} (\bibinfo{year}{1972}).

\bibitem[{\citenamefont{{Shao} and {Wex}}(2012)}]{sw12}
\bibinfo{author}{\bibfnamefont{L.}~\bibnamefont{{Shao}}} \bibnamefont{and}
  \bibinfo{author}{\bibfnamefont{N.}~\bibnamefont{{Wex}}},
  \bibinfo{journal}{Classical and Quantum Gravity}
  \textbf{\bibinfo{volume}{29}}, \bibinfo{pages}{215018}
  (\bibinfo{year}{2012}), \eprint{arXiv:1209.4503}.

\bibitem[{\citenamefont{{Shao} et~al.}(2013)\citenamefont{{Shao}, {Caballero},
  {Kramer}, {Wex}, {Champion}, and {Jessner}}}]{sck+13}
\bibinfo{author}{\bibfnamefont{L.}~\bibnamefont{{Shao}}},
  \bibinfo{author}{\bibfnamefont{R.~N.} \bibnamefont{{Caballero}}},
  \bibinfo{author}{\bibfnamefont{M.}~\bibnamefont{{Kramer}}},
  \bibinfo{author}{\bibfnamefont{N.}~\bibnamefont{{Wex}}},
  \bibinfo{author}{\bibfnamefont{D.~J.} \bibnamefont{{Champion}}},
  \bibnamefont{and}
  \bibinfo{author}{\bibfnamefont{A.}~\bibnamefont{{Jessner}}},
  \bibinfo{journal}{Classical and Quantum Gravity}
  \textbf{\bibinfo{volume}{30}}, \bibinfo{eid}{165019} (\bibinfo{year}{2013}),
  \eprint{arXiv:1307.2552}.

\bibitem[{\citenamefont{{Lazaridis} et~al.}(2009)\citenamefont{{Lazaridis},
  {Wex}, {Jessner}, {Kramer}, {Stappers}, {Janssen}, {Desvignes}, {Purver},
  {Cognard}, {Theureau} et~al.}}]{lwj+09}
\bibinfo{author}{\bibfnamefont{K.}~\bibnamefont{{Lazaridis}}},
  \bibinfo{author}{\bibfnamefont{N.}~\bibnamefont{{Wex}}},
  \bibinfo{author}{\bibfnamefont{A.}~\bibnamefont{{Jessner}}},
  \bibinfo{author}{\bibfnamefont{M.}~\bibnamefont{{Kramer}}},
  \bibinfo{author}{\bibfnamefont{B.~W.} \bibnamefont{{Stappers}}},
  \bibinfo{author}{\bibfnamefont{G.~H.} \bibnamefont{{Janssen}}},
  \bibinfo{author}{\bibfnamefont{G.}~\bibnamefont{{Desvignes}}},
  \bibinfo{author}{\bibfnamefont{M.~B.} \bibnamefont{{Purver}}},
  \bibinfo{author}{\bibfnamefont{I.}~\bibnamefont{{Cognard}}},
  \bibinfo{author}{\bibfnamefont{G.}~\bibnamefont{{Theureau}}},
  \bibnamefont{et~al.}, \bibinfo{journal}{\mnras}
  \textbf{\bibinfo{volume}{400}}, \bibinfo{pages}{805} (\bibinfo{year}{2009}),
  \eprint{arXiv:0908.0285}.

\bibitem[{\citenamefont{{Freire}
  et~al.}(2012{\natexlab{a}})\citenamefont{{Freire}, {Wex},
  {Esposito-Far{\`e}se}, {Verbiest}, {Bailes}, {Jacoby}, {Kramer}, {Stairs},
  {Antoniadis}, and {Janssen}}}]{fwe+12}
\bibinfo{author}{\bibfnamefont{P.~C.~C.} \bibnamefont{{Freire}}},
  \bibinfo{author}{\bibfnamefont{N.}~\bibnamefont{{Wex}}},
  \bibinfo{author}{\bibfnamefont{G.}~\bibnamefont{{Esposito-Far{\`e}se}}},
  \bibinfo{author}{\bibfnamefont{J.~P.~W.} \bibnamefont{{Verbiest}}},
  \bibinfo{author}{\bibfnamefont{M.}~\bibnamefont{{Bailes}}},
  \bibinfo{author}{\bibfnamefont{B.~A.} \bibnamefont{{Jacoby}}},
  \bibinfo{author}{\bibfnamefont{M.}~\bibnamefont{{Kramer}}},
  \bibinfo{author}{\bibfnamefont{I.~H.} \bibnamefont{{Stairs}}},
  \bibinfo{author}{\bibfnamefont{J.}~\bibnamefont{{Antoniadis}}},
  \bibnamefont{and} \bibinfo{author}{\bibfnamefont{G.~H.}
  \bibnamefont{{Janssen}}}, \bibinfo{journal}{\mnras}
  \textbf{\bibinfo{volume}{423}}, \bibinfo{pages}{3328}
  (\bibinfo{year}{2012}{\natexlab{a}}), \eprint{arXiv:1205.1450}.

\bibitem[{\citenamefont{{Antoniadis} et~al.}(2013)\citenamefont{{Antoniadis},
  {Freire}, {Wex}, {Tauris}, {Lynch}, {van Kerkwijk}, {Kramer}, {Bassa},
  {Dhillon}, {Driebe} et~al.}}]{afw+13}
\bibinfo{author}{\bibfnamefont{J.}~\bibnamefont{{Antoniadis}}},
  \bibinfo{author}{\bibfnamefont{P.~C.~C.} \bibnamefont{{Freire}}},
  \bibinfo{author}{\bibfnamefont{N.}~\bibnamefont{{Wex}}},
  \bibinfo{author}{\bibfnamefont{T.~M.} \bibnamefont{{Tauris}}},
  \bibinfo{author}{\bibfnamefont{R.~S.} \bibnamefont{{Lynch}}},
  \bibinfo{author}{\bibfnamefont{M.~H.} \bibnamefont{{van Kerkwijk}}},
  \bibinfo{author}{\bibfnamefont{M.}~\bibnamefont{{Kramer}}},
  \bibinfo{author}{\bibfnamefont{C.}~\bibnamefont{{Bassa}}},
  \bibinfo{author}{\bibfnamefont{V.~S.} \bibnamefont{{Dhillon}}},
  \bibinfo{author}{\bibfnamefont{T.}~\bibnamefont{{Driebe}}},
  \bibnamefont{et~al.}, \bibinfo{journal}{Science}
  \textbf{\bibinfo{volume}{340}}, \bibinfo{pages}{448} (\bibinfo{year}{2013}),
  \eprint{arXiv:1304.6875}.

\bibitem[{\citenamefont{{Callanan} et~al.}(1998)\citenamefont{{Callanan},
  {Garnavich}, and {Koester}}}]{cgk98}
\bibinfo{author}{\bibfnamefont{P.~J.} \bibnamefont{{Callanan}}},
  \bibinfo{author}{\bibfnamefont{P.~M.} \bibnamefont{{Garnavich}}},
  \bibnamefont{and}
  \bibinfo{author}{\bibfnamefont{D.}~\bibnamefont{{Koester}}},
  \bibinfo{journal}{\mnras} \textbf{\bibinfo{volume}{298}},
  \bibinfo{pages}{207} (\bibinfo{year}{1998}).

\bibitem[{\citenamefont{{Antoniadis} et~al.}(2012)\citenamefont{{Antoniadis},
  {van Kerkwijk}, {Koester}, {Freire}, {Wex}, {Tauris}, {Kramer}, and
  {Bassa}}}]{akk+12}
\bibinfo{author}{\bibfnamefont{J.}~\bibnamefont{{Antoniadis}}},
  \bibinfo{author}{\bibfnamefont{M.~H.} \bibnamefont{{van Kerkwijk}}},
  \bibinfo{author}{\bibfnamefont{D.}~\bibnamefont{{Koester}}},
  \bibinfo{author}{\bibfnamefont{P.~C.~C.} \bibnamefont{{Freire}}},
  \bibinfo{author}{\bibfnamefont{N.}~\bibnamefont{{Wex}}},
  \bibinfo{author}{\bibfnamefont{T.~M.} \bibnamefont{{Tauris}}},
  \bibinfo{author}{\bibfnamefont{M.}~\bibnamefont{{Kramer}}}, \bibnamefont{and}
  \bibinfo{author}{\bibfnamefont{C.~G.} \bibnamefont{{Bassa}}},
  \bibinfo{journal}{\mnras} \textbf{\bibinfo{volume}{423}},
  \bibinfo{pages}{3316} (\bibinfo{year}{2012}), \eprint{arXiv:1204.3948}.

\bibitem[{\citenamefont{{Stairs}}(2003)}]{sta03}
\bibinfo{author}{\bibfnamefont{I.~H.} \bibnamefont{{Stairs}}},
  \bibinfo{journal}{Living Reviews in Relativity} \textbf{\bibinfo{volume}{6}},
  \bibinfo{pages}{5} (\bibinfo{year}{2003}), \eprint{arXiv:astro-ph/0307536}.

\bibitem[{\citenamefont{{Wex}}(2014)}]{wex14}
  N. Wex, arXiv:1402.5594.

\bibitem[{\citenamefont{{Kramer} and {Wex}}(2009)}]{kw09}
\bibinfo{author}{\bibfnamefont{M.}~\bibnamefont{{Kramer}}} \bibnamefont{and}
  \bibinfo{author}{\bibfnamefont{N.}~\bibnamefont{{Wex}}},
  \bibinfo{journal}{Classical and Quantum Gravity}
  \textbf{\bibinfo{volume}{26}}, \bibinfo{eid}{073001} (\bibinfo{year}{2009}).

\bibitem[{\citenamefont{{Freire}
  et~al.}(2012{\natexlab{b}})\citenamefont{{Freire}, {Kramer}, and
  {Wex}}}]{fkw12}
\bibinfo{author}{\bibfnamefont{P.~C.~C.} \bibnamefont{{Freire}}},
  \bibinfo{author}{\bibfnamefont{M.}~\bibnamefont{{Kramer}}}, \bibnamefont{and}
  \bibinfo{author}{\bibfnamefont{N.}~\bibnamefont{{Wex}}},
  \bibinfo{journal}{Classical and Quantum Gravity}
  \textbf{\bibinfo{volume}{29}}, \bibinfo{eid}{184007}
  (\bibinfo{year}{2012}{\natexlab{b}}), \eprint{arXiv:1205.3751}.

\bibitem[{\citenamefont{{Hinshaw} et~al.}(2013)\citenamefont{{Hinshaw},
  {Larson}, {Komatsu}, {Spergel}, {Bennett}, {Dunkley}, {Nolta}, {Halpern},
  {Hill}, {Odegard} et~al.}}]{hlk+13}
\bibinfo{author}{\bibfnamefont{G.}~\bibnamefont{{Hinshaw}}},
  \bibinfo{author}{\bibfnamefont{D.}~\bibnamefont{{Larson}}},
  \bibinfo{author}{\bibfnamefont{E.}~\bibnamefont{{Komatsu}}},
  \bibinfo{author}{\bibfnamefont{D.~N.} \bibnamefont{{Spergel}}},
  \bibinfo{author}{\bibfnamefont{C.~L.} \bibnamefont{{Bennett}}},
  \bibinfo{author}{\bibfnamefont{J.}~\bibnamefont{{Dunkley}}},
  \bibinfo{author}{\bibfnamefont{M.~R.} \bibnamefont{{Nolta}}},
  \bibinfo{author}{\bibfnamefont{M.}~\bibnamefont{{Halpern}}},
  \bibinfo{author}{\bibfnamefont{R.~S.} \bibnamefont{{Hill}}},
  \bibinfo{author}{\bibfnamefont{N.}~\bibnamefont{{Odegard}}},
  \bibnamefont{et~al.}, \bibinfo{journal}{\apjs}
  \textbf{\bibinfo{volume}{208}}, \bibinfo{eid}{19} (\bibinfo{year}{2013}),
  \eprint{arXiv:1212.5226}.

\bibitem[{\citenamefont{{Damour} and {Esposito-Farese}}(1993)}]{de93}
\bibinfo{author}{\bibfnamefont{T.}~\bibnamefont{{Damour}}} \bibnamefont{and}
  \bibinfo{author}{\bibfnamefont{G.}~\bibnamefont{{Esposito-Far{\`e}se}}},
  \bibinfo{journal}{Physical Review Letters} \textbf{\bibinfo{volume}{70}},
  \bibinfo{pages}{2220} (\bibinfo{year}{1993}).

\bibitem[{\citenamefont{{Yagi} et~al.}(2014)\citenamefont{{Yagi}, {Blas},
  {Barausse}, and {Yunes}}}]{ybby14}
\bibinfo{author}{\bibfnamefont{K.}~\bibnamefont{{Yagi}}},
  \bibinfo{author}{\bibfnamefont{D.}~\bibnamefont{{Blas}}},
  \bibinfo{author}{\bibfnamefont{E.}~\bibnamefont{{Barausse}}},
  \bibnamefont{and} \bibinfo{author}{\bibfnamefont{N.}~\bibnamefont{{Yunes}}},
  \bibinfo{journal}{\prd} \textbf{\bibinfo{volume}{89}}, \bibinfo{eid}{084067}
  (\bibinfo{year}{2014}), \eprint{arXiv:1311.7144}.

\bibitem[{\citenamefont{{Stairs} et~al.}(2005)\citenamefont{{Stairs},
  {Faulkner}, {Lyne}, {Kramer}, {Lorimer}, {McLaughlin}, {Manchester}, {Hobbs},
  {Camilo}, {Possenti} et~al.}}]{sfl+05}
\bibinfo{author}{\bibfnamefont{I.~H.} \bibnamefont{{Stairs}}},
  \bibinfo{author}{\bibfnamefont{A.~J.} \bibnamefont{{Faulkner}}},
  \bibinfo{author}{\bibfnamefont{A.~G.} \bibnamefont{{Lyne}}},
  \bibinfo{author}{\bibfnamefont{M.}~\bibnamefont{{Kramer}}},
  \bibinfo{author}{\bibfnamefont{D.~R.} \bibnamefont{{Lorimer}}},
  \bibinfo{author}{\bibfnamefont{M.~A.} \bibnamefont{{McLaughlin}}},
  \bibinfo{author}{\bibfnamefont{R.~N.} \bibnamefont{{Manchester}}},
  \bibinfo{author}{\bibfnamefont{G.~B.} \bibnamefont{{Hobbs}}},
  \bibinfo{author}{\bibfnamefont{F.}~\bibnamefont{{Camilo}}},
  \bibinfo{author}{\bibfnamefont{A.}~\bibnamefont{{Possenti}}},
  \bibnamefont{et~al.}, \bibinfo{journal}{\apj} \textbf{\bibinfo{volume}{632}},
  \bibinfo{pages}{1060} (\bibinfo{year}{2005}), \eprint{arXiv:astro-ph/0506188}.

\bibitem{rcn+14} 
  B. J. Rickett, W. A. Coles, C. F. Nava,
  M. A. McLaughlin, S. M. Ransom, F. Camilo, R. D. Ferdman,
  P. C. C. Freire, M. Kramer, A. G. Lyne, and I. H. Stairs, 
  Astrophys. J. {\bf 787}, 161 (2014), arXiv:1404.1120.

\bibitem{bcs07}
  J. B. R. Battat, J. F. Chandler, and C. W. Stubbs,
  Physical Review Letters {\bf 99}, 241103 (2007), arXiv:0710.0702.  

\bibitem{mch+08} 
  H. M\"uller, S.-W. Chiow, S. Herrmann, S. Chu, and
  K.-Y. Chung, Physical Review Letters {\bf 100}, 031101 (2008), 
  arXiv:0710.3768.

\bibitem{cch+09}
  K.-Y. Chung, S.-W. Chiow, S. Herrmann, S. Chu, and H. M\"uller,
  Phys. Rev. D {\bf 80}, 016002 (2009), arXiv:0905.1929.

\bibitem{cbp+04} 
  F. Can{\` e}, D. Bear, D. F. Phillips, M. S. Rosen,
  C. L. Smallwood, R. E. Stoner, R. L. Walsworth, and V. A.
  Kosteleck{\' y}, Physical Review Letters {\bf 93}, 230801 (2004),
  arXiv:physics/0309070.

\bibitem{hac+08}
  B. R. Heckel, E. G. Adelberger, C. E. Cramer, T. S. Cook,
  S. Schlamminger, and U. Schmidt, Phys. Rev. D {\bf 78}, 092006
  (2008), arXiv:0808.2673.

\bibitem{gkv14} 
  A. H. Gomes, V. A. Kosteleck{\' y}, and
  A. J. Vargas, Phys. Rev. D {\bf 90}, 076009 (2014),
  arXiv:1407.7748. 

\end{thebibliography}
%\bibliographystyle{apsrev}
%=====================================================================

\end{document}